\documentclass{elsart}
\usepackage{latexsym}
\usepackage{amsmath}
\usepackage[dvips]{graphicx}
\newcommand{\beq}{\begin{equation}}
\newcommand{\eeq}{\end{equation}}
\newcommand{\beqa}{\begin{eqnarray}}
\newcommand{\eeqa}{\end{eqnarray}}
\newcommand{\bseq}{\begin{subequations}}
\newcommand{\eseq}{\end{subequations}}
\newcommand{\bold}{\boldsymbol}
\newcommand{\wtilde}{\widetilde}
\newcommand{\what}{\widehat}
\newcommand{\trm}{\textrm}

\begin{document}

\begin{frontmatter}

\title{Meson correlation functions in a QCD plasma}
\author{W.M. Alberico, A. Beraudo and A. Molinari}, 
\address{Dipartimento di Fisica Teorica dell'Universit\`a di Torino and \\ 
  Istituto Nazionale di Fisica Nucleare, Sezione di Torino, \\ 
  via P.Giuria 1, I-10125 Torino, Italy}


\begin{abstract}
The temporal pseudoscalar meson correlation function in a QCD plasma is investigated in a range of temperatures exceeding $T_c$ and yet of experimental interest. Only the flavour-singlet channel is considered and the imaginary time formalism is employed for the finite temperature calculations. The behaviour of the meson spectral function and of the temporal correlator is first studied in the HTL approximation, where one replaces the free thermal quark propagators with the HTL resummed ones. This procedure satisfactory describes the soft fermionic modes, but its application to the propagation of hard quarks is not reliable. An improved version of the so-called NLA scheme, which allows a better treatment of the hard fermionic modes, is then proposed. The impact of the improved NLA on the pseudoscalar temporal correlator is investigated.  
\end{abstract}
\end{frontmatter}

\section{Introduction}
The meson spectral functions in the different channels (scalar, pseudoscalar, vector, pseudovector) in the deconfined phase of QCD have been explored with various techniques, in particular with the \textit{Hard Thermal Loop} (HTL) approximation \cite{bra,lb,mus,kar}.\\
In this paper we focus most of our attention on the pseudoscalar channel where we compute the spectral function in a scheme improving upon the HTL approximation. The degrees of freedom we deal with are light quarks (massless for the sake of simplicity) and gluons and our results refer to the case of zero chemical potential (i.e. zero baryon density), a condition which is expected to be realized in the heavy ion experiments at RHIC and even better in the future experiments at LHC. The calculations are performed in the imaginary time formalism, taking at the end a proper analytical continuation to real frequencies when required.\\
To introduce the subject of our investigation we first shortly revisit the free case, where one computes the 2-point free meson correlation function and its spectral density through the convolution of two free \textit{thermal} quark propagators. Then we evaluate, for illustration in the pseudoscalar case only, the spectral function in the HTL approximation (namely using HTL resummed propagators for the fermionic lines in the loop), thus checking the results lately obtained by Karsh et al. \cite{mus,kar}.\\
Actually, for its phenomenological importance, much attention has been payed in the literature \cite{bra,lb,mus,kar,wei,critic} to the vector channel. Indeed the vector spectral function is the basic ingredient to evaluate the dilepton production rate from the plasma \cite{bra}, possibly of experimental relevance. 
However in the vector case it turns out that, in addition to the quark propagator, the quark-photon vertex should also be dressed in the HTL approximation, at variance with the pseudoscalar case where the vertex correction vanishes. We shall later shortly discuss this item.\\

One may worry that at the temperatures reached in heavy ion experiments the QCD coupling $g$ is not so small (in fact $g\sim 1$) and this could cause problems to approximation schemes, like the HTL, based on the separation of different momentum scales (i.e. $g^2T,\; gT, \; T$)\footnote{At each scale different aspects of the QGP physics disclose themselves.} which strictly holds in the weak coupling regime. Nevertheless the HTL approximation (especially in its ``improved'' versions) is able to reproduce quite well the lattice data for the thermodynamics of the QGP phase for temperatures $T\geq 2.5T_c$ as reported for example in \cite{bie}. Hence we expect such an approximation to provide reliable results for other observables as well, in the same range of temperatures.\\

Indeed the kinematical approximations which allow an analytical calculation of the quark self-energy, the basic ingredient of the HTL resummed propagator (see Appendix ~\ref{b}), are justified only if the components of the external quark four-momentum are soft, i.e. $k\sim gT$. Hence we improve upon the HTL calculation of the meson correlation function by using a prescription which allows for a better treatment of the hard fermionic modes. In this approximation, which was introduced in the analysis of the thermodynamics of the QGP \cite{blasu,blathe,blater} and is referred to as the Next to Leading Approximation (NLA), effects arising from the hard quark modes are taken into account in an effective way through a correction to the asymptotic quark thermal mass. Actually we propose a variant of the NLA which allows for a smoother transition between the two regimes of soft and hard momenta.\\

Finally, as a general remark, we observe that, within the present approximation scheme, as long as we confine ourselves to the study of the meson spectral functions (i.e. the \emph{imaginary part} of a polarization propagator) we do not need to face any problem related to renormalization.\\

Our paper is organized as follows.\\
In Sec. ~\ref{sec:def} we recall the definitions of the basic quantities we address: the meson current operator, the meson spectral function and the meson correlator along the (imaginary) temporal direction. In Sec. ~\ref{sec:free} we display the results one obtains in the free case. In Sec. ~\ref{sec:HTL}, after recalling the expression of the HTL quark spectral function, we use HTL resummed quark propagators to obtain the meson correlation function: this accounts for important effects coming from the interaction of the quarks with the thermal bath. Our calculations confirm the results previously obtained by Karsch et al. \cite{mus,kar}. In Sec. ~\ref{beyond} we examine the NLA scheme. Finally, in Sec. ~\ref{num}, we display the numerical predictions for the pseudo-scalar meson spectral function and for the temporal correlator $G(\tau)$ within the two approximation schemes (HTL and NLA) and compare them with the free case.    

\section{Thermal meson correlation function}\label{sec:def}
Adopting the same notation as in Ref. \cite{kar} we consider the following current operator, carrying the quantum numbers of a meson in the flavour-singlet channel (there is no matrix mixing the flavours):
\beq
J_M(-i\tau,\bold{x})=\bar{q}(-i\tau,\bold{x})\Gamma_M q(-i\tau,\bold{x})\;,
\eeq
where $\Gamma_M=1,\gamma^5,\gamma^{\mu},\gamma^{\mu}\gamma^5$ for the scalar, pseudoscalar, vector and pseudovector channels, respectively. We next define the fluctuation operator $\wtilde{J}_M$ as
\beq
\wtilde{J}_M(-i\tau,\bold{x})=J_M(-i\tau,\bold{x})-\langle J_M(-i\tau,\bold{x})\rangle\;,
\eeq
the average being taken on the statistical ensemble.\\
The chief quantity we address is the \textit{thermal  meson 2 point correlation function}:
\beqa
G_M(-i\tau,\bold{x}) & = & \langle \wtilde{J}_M(-i\tau,\bold{x})\wtilde{J}_M^{\dagger}(0,\bold{0})\rangle \nonumber\\
{} & = & \langle J_M(-i\tau,\bold{x})J_M^{\dagger}(0,\bold{0})\rangle-\langle J_M(-i\tau,\bold{x})\rangle\langle J_M^{\dagger}(0,\bold{0})\rangle\nonumber\\
{} & = & \frac{1}{\beta}\sum_{n=-\infty}^{+\infty}\int\frac{d^3p}{(2\pi)^3}e^{-i\omega_n\tau}e^{i\bold{p}\cdot\bold{x}}\chi_M(i\omega_n,\bold{p})\;,
\eeqa
with $\tau\in[0,\beta=1/T]$ and $\omega_n=2n\pi T$ ($n=0,\pm1,\pm2\dots$).\\
It is convenient to adopt the following spectral representation for the meson propagator in momentum space:
\beq
\chi_{M}(i\omega_n,\bold{p})=-\int_{-\infty}^{+\infty}d\omega\frac{\sigma_{M}(\omega,\bold{p})}{i\omega_n-\omega} \quad \Rightarrow \quad \sigma_{M}(\omega,\bold{p})=\frac{1}{\pi}\trm{Im}\,\chi_{M}(\omega+i\eta,\bold{p}).
\eeq
Hence it is possible to express the thermal meson propagator in a mixed representation through the \textit{spectral function} $\sigma_{M}$. Indeed starting from the definition:
\beqa
G_{M}(-i\tau,\bold{p})& = & \frac{1}{\beta}\sum_{n=-\infty}^{+\infty}e^{-i\omega_n\tau}\chi_M(i\omega_n,\bold{p}) \nonumber\\
{} & = & -\frac{1}{\beta}\sum_{n=-\infty}^{+\infty}e^{-i\omega_n\tau}\int_{-\infty}^{+\infty}d\omega\frac{\sigma_{M}(\omega,\bold{p})}{i\omega_n-\omega}\;,\nonumber 
\eeqa
and performing the sum over the Matsubara frequencies with a standard contour integration in the complex $\omega$ plane \cite{lb,bi}, one obtains \cite{kar}:
\beq
G_{M}(-i\tau,\bold{p})=\int_0^{+\infty}d\omega~\sigma_{M}(\omega,\bold{p})\frac{\cosh(\omega(\tau-\beta/2))}{\sinh(\omega\beta/2)}\;.\label{eq:gtau}
\eeq

In the following, confining ourselves to light mesons (namely the pions if one considers the pseudoscalar flavour triplet channel), we will explore in various approximation schemes the behaviour of the meson spectral function.\\
Indeed the survival of a meson as a bound state in the deconfined phase of QCD should be signaled by a delta like structure of the spectral function. We will see that this is not the case (at least according to the approaches adopted in this paper, which keep into account only the interaction of the fermionic modes with the thermal bath). On the other hand, for what concerns \textit{heavier mesons}, recent lattice results \cite{dat,asa1,asa2}, obtained in quenched QCD, do display very pronounced peaks for the charmonium spectral functions in the pseudoscalar and vector channels, which survive till $T\sim2.25T_c$. Actually, as recently pointed out by Shuryak \cite{sh,sh2}, the existence of a whole sequence of loosely bound states above $T_c$ (not necessarily colour-singlet states) would entail large cross sections among the constituents of the plasma; this in turn would justify the adoption of hydrodynamical models for the \textit{fireball} arising from the ultra-relativistic heavy-ion collisions; it would also explain certain collective phenomena, such as the elliptic flow, observed at SPS and more recently at RHIC. The QGP, at least for $T$ just above $T_c$, seems to behave more like a liquid with a very small viscosity rather than like a gas of quasiparticles.\\
In the light of these results, which indicate the survival of mesonic resonances in the QGP phase, the existence of bound states in the broken-symmetry phase of the three-dimensional Ising model \cite{cas1,cas2} gains considerable interest. We remind the reader that, on the lattice, the 3d Ising model is related through a duality transformation to the $Z_2$ gauge model \cite{sav}, the center of the group $SU(2)$.   
  
\section{Free spectral functions}\label{sec:free}

\begin{figure}[tp]
\begin{center}
\includegraphics[clip,width=0.3\textwidth]{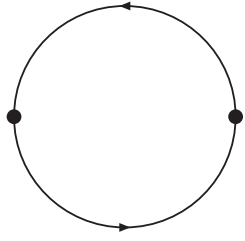}
\caption{Diagram illustrating the free meson correlation function. The propagation lines are free thermal quark propagators. The two interaction vertices correspond to $\Gamma$ matrices.}\label{freemeson} 
\end{center}
\end{figure}

We start by revisiting the computation of the free meson spectral functions for the different channels. We use the short-hand notation $(\tau,\bold{x})$ for $(t=-i\tau,\bold{x})$. In order to display more transparently how one gets the final result, we write explicitly the spinor indices. We have to compute the expression
{\setlength\arraycolsep{1pt}
\beqa
\langle J_M(\tau,\bold{x})J_M^{\dagger}(0,\bold{0})\rangle & = & \langle\bar{q}(\tau,\bold{x})\Gamma_M q(\tau,\bold{x})(\bar{q}(0,\bold{0})\Gamma_M q(0,\bold{0}))^\dagger\rangle \nonumber\\
{} & = & \langle\bar{q}_{\alpha}(\tau,\bold{x})q_{\beta}(\tau,\bold{x})\bar{q}_{\lambda}(0,\bold{0})q_{\mu}(0,\bold{0})\rangle\times\nonumber\\
{} & {} & \times(\Gamma_M)_{\alpha\beta}(\gamma^0\Gamma_M^{\dagger}\gamma^0)_{\lambda\mu}\
\eeqa}
diagrammatically displayed, after subtracting the disconnected contribution, in Fig. (\ref{freemeson}).\\
Since the field operators inside the average over the ensemble are already \textit{time-ordered} (being $\tau>0$) we apply Wick's theorem. From the customary definitions (see Appendix ~\ref{a}):
\beqa
\langle T_\tau\,[q_\beta(\tau,\bold{x})\bar{q}_{\lambda}(0,\bold{0})]\rangle & = & S_{\beta\lambda}^F(\tau,\bold{x})\\
\langle T_\tau\,[\bar{q}_\alpha(\tau,\bold{x})q_{\mu}(0,\bold{0})]\rangle & = & -S_{\mu\alpha}^F(-\tau,-\bold{x}) \quad \nonumber\\
{} & = & S_{\mu\alpha}^F(\beta-\tau,-\bold{x})\;,
\eeqa
$S^F$ being the quark propagator, it follows that:
\beq
\langle\wtilde{J}_M(\tau,\bold{x})\wtilde{J}_M^{\dagger}(0,\bold{0})\rangle =\trm{Tr}(\Gamma_M S_F(\tau,\bold{x})\gamma^0\Gamma_M^{\dagger}\gamma^0 S_F(\beta-\tau,-\bold{x}))\;.
\eeq
Hence in Fourier space the mesonic 2 point correlation function reads \cite{kar}:
\beq
\chi_{M}(i\omega_l,\bold{p})\!=\!-2N_c\!\frac{1}{\beta}\!\sum_{n=-\infty}^{+\infty}\!\!\int\!\frac{d^3k}{(2\pi)^3}\trm{Tr}[\Gamma_{M}S_{F}(i\omega_n,\bold{k})\gamma^0\Gamma_{M}^{\dagger}\gamma^{0}S_{F}(i\omega_n-i\omega_l,\bold{k}-\bold{p})]\label{generale} ,
\eeq
where the mesonic frequency is $\omega_{l}=2l\pi T$, while in the quark propagators $\omega_n=\textrm{(2n+1)}\pi T$, $l$ and $n$ assuming any integer value. The overall factor $2N_c$ ($N_c$ being the number of colours) comes from the trace over the light flavours and colours.\\
If one considers the flavour triplet channel the overall flavour factor $N_f=2$ has to be replaced by
\beq
\trm{Tr}\left[\tau^a\tau^b\right]=2\delta^{ab}\;,
\eeq
where $\tau^a$ and $\tau^b$ are Pauli matrices acting in the flavour space. The same prescription allows one to translate the results obtained in the following for the interacting case to the non-singlet channel.\\

We can now write down explicitly the free meson propagator in the different channels getting:
{\setlength\arraycolsep{1pt}
\beqa
\chi^\trm{s}(i\omega_l,\bold{p}) & = & -2N_c\!\frac{1}{\beta}\!\sum_{n=-\infty}^{+\infty}\!\!\int\!\frac{d^3k}{(2\pi)^3}\trm{Tr}[S_{F}(i\omega_n,\bold{k})S_{F}(i\omega_n\!-i\omega_l,\bold{k}-\bold{p})]\\
\chi^{\trm{ps}}(i\omega_l,\bold{p}) & = & 2N_c\!\frac{1}{\beta}\!\sum_{n=-\infty}^{+\infty}\!\!\int\!\frac{d^3k}{(2\pi)^3}\trm{Tr}[\gamma^5 S_{F}(i\omega_n,\bold{k})\gamma^5 S_{F}(i\omega_n\!-i\omega_l,\bold{k}-\bold{p})]\label{ps} \\
\chi^{\trm{v}}(i\omega_l,\bold{p}) & = & -2N_c\!\frac{1}{\beta}\!\sum_{n=-\infty}^{+\infty}\!\!\int\!\frac{d^3k}{(2\pi)^3}\trm{Tr}[\gamma^\mu S_{F}(i\omega_n,\bold{k})\gamma_\mu S_{F}(i\omega_n\!-i\omega_l,\bold{k}\!-\bold{p})]\label{vector} \\
\chi^{\trm{pv}}(i\omega_l,\bold{p}) & = & 2N_c\!\frac{1}{\beta}\!\sum_{n=-\infty}^{+\infty}\!\!\int\!\frac{d^3k}{(2\pi)^3}\trm{Tr}[\gamma^\mu\gamma^5 S_{F}(i\omega_n,\bold{k})\gamma^5\gamma_\mu S_{F}(i\omega_n\!-i\omega_l,\bold{k}\!-\bold{p})]\;,\nonumber\\ 
\eeqa}
where in the vector (see also Ref. \cite{kar}) and pseudovector channels the Dirac matrices have been contracted with the Minkowskian metric tensor $g_{\mu\nu}$. Note that in the massless, non interacting case, since $\{S_F,\gamma^5\}=0$, $\chi^\trm{s}\!=\!\chi^{\trm{ps}}$ and $\chi^{\trm{v}}\!=\!\chi^{\trm{pv}}$.\\
Next we evaluate the imaginary part of the above quantities, thus obtaining the free meson spectral function. We sketch the calculation in the vector case and just quote the results for the other channels. To simplify the problem we confine ourselves to the $\bold{p}=0$ limit. Yet, even in this limit, we can grasp important aspects of the plasma physics once the interaction of the quarks with the thermal bath is switched on. Indeed $\sigma(\omega,0)$ (the zero momentum projection of the spectral function) is the quantity which is studied on the lattice in order to determine whether a meson can survive as a bound state also in the deconfined phase of QCD \cite{dat}. Furthermore, as previously mentioned, $\sigma^{\trm{v}}(\omega,0)$ is directly related to the back-to-back dilepton production rate \cite{bra}.\\
Introducing the four-vectors $A=(i\omega_{n},\bold{k})$ and $B=(i\omega_{n}-i\omega_{l},\bold{k})$ one can easily perform the trace in Eq. (\ref{vector}) obtaining:
\beq
\trm{Tr}[\gamma^\mu(A\hspace{-.20cm}{\slash}+m)\gamma_\mu(B\hspace{-.20cm}{\slash}+m)]=-8(A\cdot B)+16m^2\;,
\eeq
$m$ being the bare quark mass.   
Thus what is left out to compute is the following sum over the Matsubara frequencies
\beq\label{somma}
\frac{1}{\beta}\sum_{n}\frac{8(\omega_n^2+E_k^2)-8\omega_n\omega_l+8m^2}{(\omega_n^2+E_k^2)[(\omega_n-\omega_l)^2+E_k^2]}\;,
\eeq
which is easily done with the help of the formulas quoted in Appendix ~\ref{c}. Then, after performing a proper analytical continuation letting $i\omega_l\rightarrow \omega +i\eta^+$ ($\omega$ being real), one takes the imaginary part of the obtained result. Note that the term in round brackets in the numerator of Eq. (\ref{somma}) leads to a purely real result (which, incidentally, would require to be renormalized) and hence does not contribute. We thus recover the well-known result for the free vector spectral function (for $\omega>0$):
{\setlength\arraycolsep{1pt}
\beqa
\sigma^\trm{v}(\omega,\bold{0}) & = & -2N_c\int\frac{d^3k}{(2\pi)^3}(1-\tilde{n}(\omega/2))\delta(\omega-2E_k)\cdot 2\cdot\left(\frac{m^2}{E_k^2}+\frac{\omega}{E_k}\right)\nonumber\\
{} & = & \frac{N_c}{4\pi^2}\theta(\omega-2m)\sqrt{1-\left(\frac{2m}{\omega}\right)^2}\omega^2\tanh(\omega/4T)\left(-2-\left(\frac{2m}{\omega}\right)^2\right)\;.\nonumber\\
\eeqa} 
In the above one recognizes the typical threshold behaviour, the square root stemming from the integration over the phase space.\\
Globally the results for the different channels can be summarized in the compact expression:
\beq
\sigma_{M}^{\textrm{free}}(\omega,\bold{0})=\frac{N_c}{4\pi^2}\theta(\omega-2m)\sqrt{1-\left(\frac{2m}{\omega}\right)^2}\omega^2\tanh(\omega/4T)\left(a+b\left(\frac{2m}{\omega}\right)^2\right)\; ,
\eeq
where the coefficient $(a,b)$ turn out to be given respectively by (1,-1), (1,0), (-2,-1), (-2,3) in the scalar, pseudoscalar, vector and pseudovector channel\footnote{These coefficients do not coincide with the ones of Refs. \cite{mus,kar}, due to our choice of working with ordinary Dirac matrices satisfying the anti-commutation relation $\{\gamma^\mu,\gamma^\nu\}=2g^{\mu\nu}$, $g^{\mu\nu}$ being the Minkowskian metric tensor.}.
\begin{figure}[htp]
\begin{center}
\includegraphics[clip,width=0.7\textwidth]{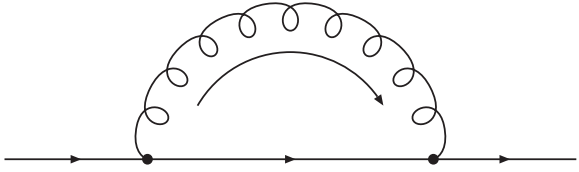}
\caption{The diagram contributing to the quark self-energy in the HTL approximation. The wavy line denotes a gluon. The external quark lines are soft.}\label{quarkselfhtl} 
\end{center}
\end{figure}

\begin{figure}[htp]
\begin{center}
\includegraphics[clip,width=0.7\textwidth]{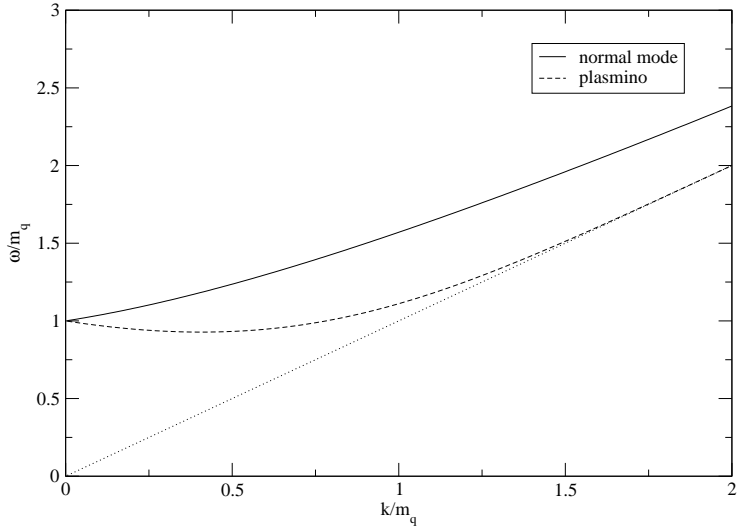}
\caption{Dispersion relations $\omega=\omega_\pm(k)$ corresponding to the quasiparticle poles of the HTL fermion propagator. Dimensionless units $k/m_q$ and $\omega/m_q$ are used, $m_q$ being the thermal quark gap mass (see text). The normal (continuous line) and the plasmino (dashed line) modes are displayed together with the light-cone (dotted line).}\label{fermdisp} 
\end{center}
\end{figure}

\begin{figure}[!htp]
\begin{center}
\includegraphics[clip,width=0.7\textwidth]{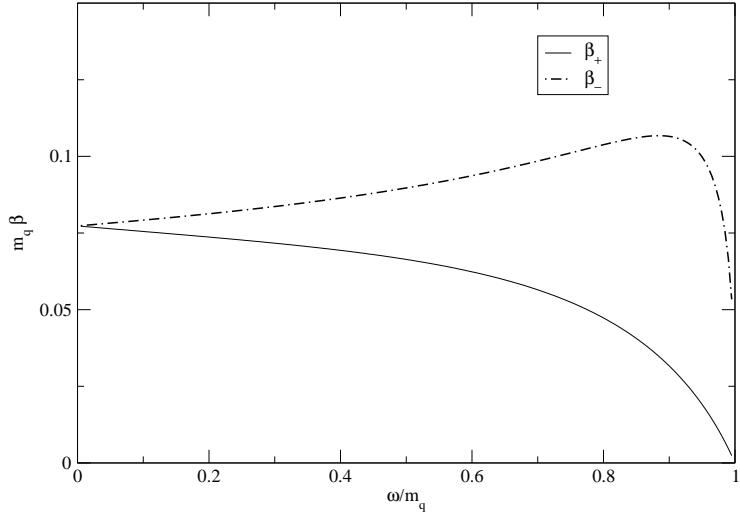}
\caption{Dimensionless (continuum) spectral function $m_q\cdot\beta_\pm(\omega,k)$ for space-like momenta at $k=m_q$ as a function of $\omega/m_q$. The maximum of $\beta_{-}$ stems from the second zero of the function $\trm{Re}(\,^\star\Delta_{-}^{-1})$, occurring in the space-like region. Owing to the large imaginary part of the self-energy, this maximum does not show up as a well pronounced peak corresponding to a quasi-particle excitation.}\label{Landau} 
\end{center}
\end{figure}

\section{Spectral functions in HTL approximation}\label{sec:HTL}
In this section we evaluate the meson spectral function in HTL approximation for the pseudoscalar and vector channel. In both cases this amounts to dressing the quark propagator with the self-energy of Fig. \ref{quarkselfhtl}, keeping only the term proportional to $g^2T^2$, as discussed in detail, for example, in \cite{lb}. In performing this task it is useful to express the fermionic propagator thus obtained through its spectral function (see (\ref{spectral})). The latter, in HTL, turns out to read \cite{bra,lb}:
\beq
\rho_{\pm}(\omega,k)=\frac{\omega^2-k^2}{2m_q^2}[\delta(\omega-\omega_{\pm})+\delta(\omega+\omega_{\mp})]+\beta_{\pm}(\omega,k)\theta(k^2-\omega^2)\;,\label{2part}
\eeq
with
\beq
\beta_{\pm}(\omega,k)=-\frac{m_q^2}{2}\frac{\pm\omega-k}{\left[k(-\omega\pm k)+m_q^2\left(\pm1-\frac{\pm\omega-k}{2k}\ln\frac{k+\omega}{k-\omega}\right)\right]^2+\left[\frac{\pi}{2}m_q^2\frac{\pm\omega-k}{k}\right]^2}\;,
\eeq
where the two dispersion relations $\omega_\pm(k)$ are shown in Fig. (\ref{fermdisp}), while the behaviour of $\beta_\pm(\omega,k)$ is plotted in Fig. (\ref{Landau}). In the above $m_q=gT/\sqrt{6}$ is the so-called \emph{thermal gap mass} of the quark, clearly appearing in Fig. (\ref{fermdisp}) at $k=0$. The HTL spectral functions in Eq. (\ref{2part}) consist of two pieces: a pole term, arising from the zeros of the real part of the denominators of $^{\star}\Delta_{\pm}$ in Eq. (\ref{delta}), and a continuum term, corresponding to the Landau damping of a quark propagating in the bath.
In the time-like domain $\omega>k$ the dispersion relation $\omega_{+}(k)$ solves the equation $^{\star}\Delta_{+}^{-1}(\omega_{+}(k),k)=0$ and corresponds to the propagation of a quasi-particle with chirality and helicity eigenvalues of the same sign. On the other hand $\omega_{-}(k)$ satisfies the equation $^{\star}\Delta_{-}^{-1}(\omega_{-}(k),k)=0$ and describes the propagation of an excitation, referred to as \textit{plasmino}, with negative helicity over chirality ratio. Note that both these excitations are undamped at this level of approximation, since the logarithm in $^{\star}\Delta_{\pm}$ doesn't develop any imaginary part in the time-like domain. On the contrary, for space-like momenta $\omega<k$, the quark spectral function receives contribution from the imaginary part stemming from the logarithm contained in Eq. (\ref{den}): this gives rise to the term $\beta_{\pm}$ in Eq. (\ref{2part}).\\
We now address the meson spectral functions in two specific cases.

\subsection{Pseudoscalar channel}

\begin{figure}[tp]
\begin{center}
\includegraphics[clip,width=0.3\textwidth]{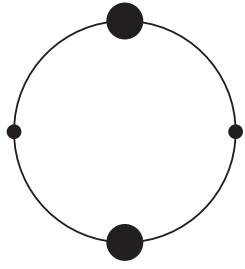}
\caption{Diagram giving the pseudoscalar meson 2-point function in the HTL approximation. The interaction vertices are $\gamma^5$ matrices, while the dressed lines are HTL resummed propagators. No vertex correction is required within this scheme.}\label{meson2} 
\end{center}
\end{figure}

The pseudoscalar vertex receives no HTL correction as shown in \cite{tho}. Hence in the pseudoscalar channel, following Ref. \cite{kar}, one has simply to replace, in Eq.(\ref{ps}), the free thermal quark propagators with HTL resummed ones (which we shall denote with a star), 
obtaining:
\beq
\chi^{\trm{ps}}(i\omega_l,\bold{p})=2N_c\!\frac{1}{\beta}\!\sum_{n=-\infty}^{+\infty}\!\!\int\!\frac{d^3k}{(2\pi)^3}\trm{Tr}[\gamma^{5}\, ^{\star}S(i\omega_n,\bold{k})\gamma^{5}\, ^{\star}S(i\omega_n\!-i\omega_l,\bold{k}-\bold{p})]\; .
\eeq
Setting $\bold{q}=\bold{k}-\bold{p}$ and making use of the spectral representation (\ref{fermspec}) of the quark propagator one gets:
\begin{multline}
\chi^{\trm{ps}}(i\omega_l,\bold{p})\!=\!2N_c\frac{1}{\beta}\!\sum_{n=-\infty}^{+\infty}\!\!\int\!\frac{d^3k}{(2\pi)^3}\int_{-\infty}^{+\infty}d\omega_1\int_{-\infty}^{+\infty}d\omega_2\frac{1}{i\omega_n-\omega_1}\frac{1}{i\omega_n-i\omega_l-\omega_2}\times\\
\times\trm{Tr}[\gamma^5\rho_{\trm{HTL}}(\omega_1,\bold{k})\gamma^5\rho_{\trm{HTL}}(\omega_2,\bold{q})]\;.
\end{multline}
Summing over the Matsubara frequencies with a standard contour integration, performing the usual analytical continuation $i\omega_{l}\rightarrow\omega +i\eta^+$ and taking the imaginary part of the result thus obtained one gets:
\begin{multline}
\sigma^{\trm{ps}}(\omega,\bold{p})=-2N_c\int\!\frac{d^3k}{(2\pi)^3}(e^{\beta\omega}-1)\int_{-\infty}^{+\infty}d\omega_1\int_{-\infty}^{+\infty}d\omega_2\tilde{n}(\omega_1)\tilde{n}(\omega_2)\times\\
\times\delta(\omega-\omega_1-\omega_2)\cdot\trm{Tr}[\gamma^5\rho_{\trm{HTL}}(\omega_1,\bold{k})\gamma^5\rho_{\trm{HTL}}(-\omega_2,\bold{q})]\;.\label{eq:sigmap}
\end{multline}
Now, inserting Eq. (\ref{spectral}) into Eq. (\ref{eq:sigmap}) and since
\bseq
\begin{align}
\trm{Tr}\left[\gamma^5\frac{\gamma^0\mp\bold{\gamma\cdot\hat{k}}}{2}\gamma^5\frac{\gamma^0\mp\bold{\gamma\cdot\hat{q}}}{2}\right]&=-(1-\bold{\hat{k}\cdot\hat{q}}),\\
\trm{Tr}\left[\gamma^5\frac{\gamma^0\mp\bold{\gamma\cdot\hat{k}}}{2}\gamma^5\frac{\gamma^0\pm\bold{\gamma\cdot\hat{q}}}{2}\right]&=-(1+\bold{\hat{k}\cdot\hat{q}})\;,
\end{align}
\eseq
one gets \cite{kar}:
\begin{multline}
\sigma^{\trm{ps}}(\omega,\bold{p})\!=\!2N_c\int\!\frac{d^3k}{(2\pi)^3}(e^{\beta\omega}-1)\int_{-\infty}^{+\infty}d\omega_1\int_{-\infty}^{+\infty}d\omega_2\tilde{n}(\omega_1)\tilde{n}(\omega_2)\delta(\omega-\!\omega_1-\!\omega_2)\times\\
\times\left\{(1+\bold{\hat{k}\cdot\hat{q}})[\rho_+(\omega_1,k)\rho_+(\omega_2,q)+\rho_-(\omega_1,k)\rho_-(\omega_2,q)]+\right.\\
+\left.(1-\bold{\hat{k}\cdot\hat{q}})[\rho_+(\omega_1,k)\rho_-(\omega_2,q)+\rho_-(\omega_1,k)\rho_+(\omega_2,q)]\right\}\;.\label{eq:sigmapcompl}
\end{multline}
where the identity $\rho_{+}(-\omega,k)=\rho_{-}(\omega,k)$ has been used.\\
In the case $\bold{p}=0$ the above formula reduces to:
\beqa
\sigma^{\trm{ps}}(\omega,\bold{0}) & = & \frac{2N_c}{\pi^2}(e^{\beta\omega}-1)\int_0^{+\infty}dk\, k^2\int_{-\infty}^{+\infty}d\omega_1\int_{-\infty}^{+\infty}d\omega_2 \tilde{n}(\omega_1)\tilde{n}(\omega_2)\nonumber\\
{} & {} & \delta(\omega-\omega_1-\omega_2)[\rho_{+}(\omega_1,k)\rho_{+}(\omega_2,k)+\rho_{-}(\omega_1,k)\rho_{-}(\omega_2,k)]\;.\nonumber\\ \label{pseudo}
\eeqa
Inserting then into Eq. (\ref{pseudo}) the explicit expression for $\rho_{\pm}$ given in Eq. (\ref{2part}), one finds, as first pointed out in Ref. \cite{bra}, that the meson spectral function consists of the sum of three terms: pole-pole (pp), pole-cut (pc) and cut-cut (cc). These different contributions will be discussed in Sec. ~\ref{num}.   

\subsection{Vector channel}
\begin{figure}[htp]
\begin{center}
\includegraphics[clip,width=0.6\textwidth]{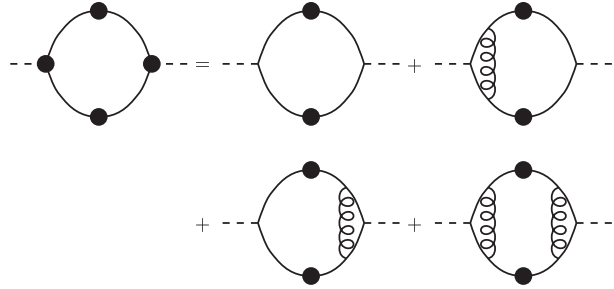}
\caption{The diagrams contributing to the vector meson spectral function in the HTL approximation. Dressed lines and vertices denote HTL resummed propagators and HTL effective quark-photon vertices respectively. The momentum flowing in the loops dressing the vertices is hard. The external dashed lines denote a photon (or a probe with the same quantum-numbers).}\label{fig:vector} 
\end{center}
\end{figure}
The vector meson spectral function in the deconfined phase, being related to the thermal dilepton production rate from the plasma, is clearly of importance for the QGP phenomenology. When computed in the HTL approximation it should provide a reliable estimate for the soft dilepton production rate. For the sake of completeness we quote here the HTL vector spectral function previously obtained in \cite{bra,lb,mus,kar}, whose meaning in term of Feynman diagrams is illustrated in Fig. (\ref{fig:vector}). Its expression is considerably more cumbersome than the pseudoscalar one reading \cite{bra,lb,mus,kar}
\begin{multline}
\sigma^\trm{v}(\omega,\bold{0})=\frac{4N_c}{\pi^2}(e^{\beta\omega}-1)\int_0^{+\infty}\!dk\, k^2\int_{-\infty}^{+\infty}\!d\omega_1\int_{-\infty}^{+\infty}\!d\omega_2 \tilde{n}(\omega_1)\tilde{n}(\omega_2)\delta(\omega-\omega_1-\omega_2)\\
\times\left\{4\left(1-\frac{\omega_1^2-\omega_2^2}{2k\omega}\right)^2\rho_{+}(\omega_1,k)\rho_{-}(\omega_2,k)+\left(1+\frac{\omega_1^2+\omega_2^2-2k^2-2m_q^2}{2k\omega}\right)^2\right.\\
\times\rho_{+}(\omega_1,k)\rho_{+}(\omega_2,k)+\left(1-\frac{\omega_1^2+\omega_2^2-2k^2-2m_q^2}{2k\omega}\right)^2\rho_{-}(\omega_1,k)\rho_{-}(\omega_2,k)\\
\left. +\Theta(k^2-\omega_1^2)\frac{m_q^2}{4k\omega^2}(1-\frac{\omega_1^2}{k^2})\left[ (1+\frac{\omega_1}{k})\rho_{+}(\omega_2,k)+(1-\frac{\omega_1}{k})\rho_{-}(\omega_2,k)\right] \right\}\;\label{eq:vector}
\end{multline}
and is derived by dressing in Eq. (\ref{vector}) both the propagation lines (replacing the free thermal quark propagators with HTL resummed ones) and the bare vertices. For the latter the following effective HTL vertex for the coupling of a quark with a vector probe is adopted:
\beq
^{\star}\Gamma_{\mu}(P_1,P_2)=\gamma_\mu+m_q^2\int\frac{d\Omega_k}{4\pi}\frac{\what{K}\hspace{-.25cm}{\slash}\;\what{K}_{\mu}}{\left[(P_1\cdot\what{K})(P_2\cdot\what{K})\right]}\;,\label{eq:vertex}
\eeq
where $\what{K}\!=\!(1,\bold{\hat{k}})$, $P_1\!=\!(i\omega_{n_1},\bold{p_1})$ is the four-momentum of the incoming quark, while $P_2\!=\!(i\omega_{n_2},\bold{p_2})$ refers to the outgoing one. Again we note that, because our $\gamma^\mu$ matrices satisfy the usual Dirac algebra (with the Minkowskian metric tensor $g^{\mu\nu}$), we get a different sign in front of the vertex correction with respect to what quoted in Ref. \cite{lb}.\\

At this point a careful analysis of the momenta over which one performs the integration is appropriate. In fact, according to the  original proposal by Braaten and Pisarski \cite{pis} for constructing an effective theory for the soft modes in the QGP, one has to replace every soft fermionic line (i.e. with $K\sim gT$) with an HTL resummed propagator, while the hard propagation lines should be left undressed. This is because, for $K\sim gT$, also the self-energy $\Sigma_{\trm{HTL}}(K)\sim gT$: hence all the diagrams with repeated self-energy insertions in the fermionic lines give contributions of the same order, i.e.
\begin{displaymath}
\frac{-1}{K\hspace{-.25cm}{\slash}}\cdot(-\Sigma)\frac{-1}{K\hspace{-.25cm}{\slash}}\cdot(-\Sigma)\frac{-1}{K\hspace{-.25cm}}\cdot\dots\sim\frac{-1}{K\hspace{-.25cm}{\slash}}\sim\frac{1}{gT}\;.
\end{displaymath}
In addition, if all the lines entering into a vertex are soft one has to replace it with the HTL one given by Eq. (\ref{eq:vertex}), otherwise the vertex should be left untouched. On the other hand a hard fermion propagator (i.e. with $K\sim T$) can be expanded perturbatively, since the kinetic term dominates over the interaction, and in first approximation one can keep its non-interacting expression. Hence the integration over $k$ in Eq. (\ref{eq:vector}) really should cover only the soft regime of momenta, while in the integration over the hard momenta domain one should use the bare quantities.\\  

Actually the vector meson spectral function in the deconfined phase and the associated thermal dilepton rate have been recently evaluated on the lattice \cite{bile1,bile2,bile3}. It turns out that their behaviour at low energies is very different from the HTL result given by Eq. (\ref{eq:vector}). Specifically, whereas the HTL spectral function develops an infrared divergence for $\omega\to0$ (stemming from the cut-cut contribution) \cite{bra}, the lattice result drops to zero below $\omega/T\simeq3$, suggesting a threshold effect due to a thermal mass acquired by the quarks. In this connection we remind the reader that the spectral function of a meson is not measured directly on the lattice, but has to be obtained by inverting Eq. (\ref{eq:gtau}). A procedure referred to as Maximum Entropy Method \cite{dat,asa1,asa2,bile1,bile2,bile3} is usually employed to accomplish this task.\\
   
Finally, for what concern the experimental results, in a heavy ion collision dilepton pairs are produced also in processes different from the annihilation of $q\bar{q}$ quasi-particles in the plasma phase\footnote{Actually, even in the deconfined phase as we will discuss in detail for the pseudoscalar case, the HTL meson spectral function gets contributions from far from trivial many-body processes.}. To get an estimate for the production rate which can be eventually compared with the experiment, one has to account also for other sources \cite{wei}: in particular the (non-thermal) contribution due to Drell-Yan processes, the dileptons coming from the hadronic phase and the decay of vector mesons after the freeze-out.\\

In the following we shall concentrate on the pseudoscalar channel only, aiming at an improved treatment of the latter.\\ 

\section{Beyond HTL}\label{beyond}
In this section we attempt to improve upon the quark spectral functions appearing in the loop of Fig. (\ref{meson2}). For this purpose we observe again that, in making the convolution of the two fermionic propagators, one has to integrate over \textit{all the scale of momenta} (hard and soft). Now, while the HTL approximation is supposed to dress correctly the propagation of the \textit{soft modes}, this is not so for what concerns the \textit{hard modes}. Thus, by replacing naively the free thermal quark propagators in Eq. (\ref{generale}) with HTL resummed ones, one treats incorrectly the contribution to the integral arising from hard momenta. This is analogous to what happens when one tries to evaluate thermodynamical quantities for the QGP, like the entropy and the baryon density, in a pure HTL approximation. As pointed out in Ref. \cite{bie}, in HTL one gets the right contribution of order $g^2$ to such quantities, but \textit{only} part of the $g^3$ term (actually this one is strictly a non-perturbative contribution, being non analytical in $\alpha_s=g^2/4\pi$), namely the contribution arising from the soft modes.\\        

Now we shortly examine how the Next to Leading contributions, amounting to dress properly the propagation of the hard degrees of freedom in a QGP, are kept into account in a self-consistent calculation of quantities like the entropy, the baryon density and the quark-number susceptibilities \cite{bie,blasu,blater}.\\
\begin{figure}[tp]
\begin{center}
\includegraphics[clip]{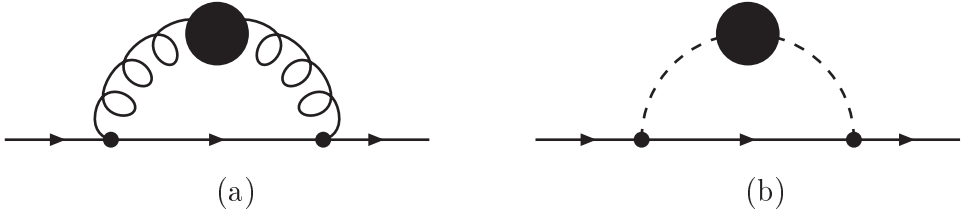}
\caption{Next to Leading corrections to a hard quark propagator, arising from the interaction with a soft transverse (a) and longitudinal (b) gluon: for these HTL resummed propagators are used. Thus the two diagrams receive contributions from all orders in perturbation theory.}\label{quarkselfnlo} 
\end{center}
\end{figure}
These observables receive a non-analytical contribution of order $g^3$ both from the soft modes and from the proper dressing of the hard propagators. To get the latter one has to evaluate the correction $\delta\Sigma_{\pm}(\omega=\pm k)$\footnote{Note that only the light-cone projection of the hard quark self-energy enters into the thermodynamical calculations. The same is true for the hard transverse gluon as well. Since on the light-cone the HTL self-energies coincide with the full one-loop calculation, the entropy and the baryon density might be privileged quantities to study within the HTL approximation.} to the self energy of a normal quark mode (with hard momentum). Such a contribution corresponds to the two diagrams displayed in Fig. \ref{quarkselfnlo} where the hard quark interacts with a \textit{soft} gluon, longitudinal and transverse, respectively, whose \textit{resummed} propagators are the HTL ones (being the gluons soft such an approximation is fully justified). Note that this is a non-perturbative procedure, the two diagrams in Fig. \ref{quarkselfnlo} getting contributions from all orders of perturbation theory.\\
Actually the correction $\delta\Sigma_{\pm}(\omega=\pm k)$ is a far from trivial function of the momentum and generally it can be dealt with only numerically. Nevertheless its contribution of order $g^3$ to the entropy and the baryon density can be analytically evaluated. Indeed in \cite{bie,blasu,blater} the authors replace this momentum-dependent self-energy correction with a constant shift of the asymptotic quark mass in order to reproduce the right coefficients of the terms of order $g^3$ coming from the hard sector for the entropy and the baryon density\footnote{Analogously the correction $\delta\Pi_T(\omega=\pm k)$ to the self-energy of a hard momentum transverse gluon is absorbed in a shift of its asymptotic mass.}.\\
This mass correction turns out to be negative, reading \cite{bie,blasu,blater}:
\beq
\delta m_{\infty}^2=-\frac{1}{2\pi}\,g^2\,C_f\,T\,\hat{m}_D\;,\label{correzione}
\eeq   
where (in the case of zero chemical potential) the Debye screening mass in the HTL approximation is:
\beq
\hat{m}_D=\sqrt{\frac{N_c}{3}+\frac{N_f}{6}}\,gT\;,
\eeq
being $N_f$ the number of flavours and $C_f=4/3$ the \emph{quadratic Casimir operator} for the fundamental representation of $SU(3)$.\\
It is worth stressing that at the experimentally accessible temperatures (i.e. a few times $T_c$) $g\ge 1$, hence the (negative) correction (\ref{correzione}) becomes of the same order of magnitude of the HTL asymptotic mass $\hat{m}_\infty$ which reads
\beq
\hat{m}_{\infty}^2=2m_q^2=\frac{g^2T^2}{3}\;,\label{eq:htlinf}
\eeq
thus possibly leading to a tachyonic singularity.\\
This problem was addressed in Ref. \cite{bie} where two possible solutions where proposed. The first amounts to replace the expression $m_{\infty}^2=\hat{m}_{\infty}^2+\delta m_{\infty}^2$ with its first Pade' approximant, namely
\beq
m_{\infty}^2=\hat{m}_{\infty}^2+\delta m_{\infty}^2\quad\longrightarrow\quad m_{\infty}^2=\frac{\hat{m}_{\infty}^2}{\displaystyle{1-\frac{\delta m_{\infty}^2}{\hat{m}_{\infty}^2}}}\label{massapade'}\;.
\eeq

The other starts with the correct equation for the Next to Leading quark asymptotic mass $m_\infty$, namely
\beq
m_{\infty}^2=\hat{m}_{\infty}^2-\frac{1}{2\pi}\,g^2\,C_f\,\sqrt{N_c+\frac{N_f}{2}}\,T\,\hat{m}_{\infty}\;,
\eeq
and replaces $\hat{m}_{\infty}\rightarrow m_{\infty}$ in the term linear in $\hat{m}_\infty$, thus obtaining the ``self-consistent like'' equation:
\beq
m_{\infty}^2=\hat{m}_{\infty}^2-\frac{1}{2\pi}\,g^2\,C_f\,\sqrt{N_c+\frac{N_f}{2}}\,T\,m_{\infty}\label{massanla}\;,
\eeq 
whose positive root provides the asymptotic mass for the hard quark. In the following, in accord with the literature \cite{bie}, we shall refer to such an approach as the Next to Leading Approximation (NLA).\\
To replace in the term linear in $\hat{m}_\infty$ the HTL mass with its ``self-consistent'' value may appear somewhat arbitrary. Yet the effects of such a procedure were tested in a scalar toy model \cite{bie}, where one can solve the gap equation exactly (although numerically), and the discrepancy between the exact result and the above discussed procedure turned out to be $\sim1\%$ even for large values of the coupling ($g\sim2$), at least as far as the entropy is concerned.\\
The NLA asymptotic mass $m_\infty$ applies to the hard quark modes. Since the soft modes are correctly described by the HTL approximation, the strategy followed in Refs. \cite{bie,blasu,blater} was to single them out by introducing a cutoff $\Lambda$ at an intermediate scale of momenta. A good choice for the latter is:
\beq
\Lambda=\sqrt{2\pi T\hat{m}_D}\quad,\label{spatialcutoff}
\eeq 
which represents the geometric mean of the spacing between the Matsubara frequencies (hard scale) and the Debye screening mass in HTL (soft scale). Later we will investigate the sensitivity of the results to the value of this cutoff.\\
Now, in accord with the above discussion, in the integration over the spatial momenta in the formulas for the entropy and the baryon density one is allowed to keep the HTL expression of the integrand for $k<\Lambda$ ($k$ being the modulus of the three-momentum). Hence the entropy will receive contribution from the quasi-particle poles (normal quark mode, plasmino, transverse and longitudinal gluons) and from the Landau damping as well. For $k>\Lambda$ instead one takes into account only the propagation of the \textit{physical modes} (normal quarks and transverse gluons), which are now dressed by the asymptotic masses obtained in NLA, neglecting the damping contribution. A comparisons of the results obtained in such a framework with the one coming from lattice calculations are reported in Refs. \cite{bie,blasu,blater} and the agreement appears satisfactory.\\

A similar strategy is here applied to the evaluation of mesonic correlation functions. In particular we explore in detail how an improved approximation at the level of the quark propagator is reflected on the (zero momentum) meson spectral function (in the pseudo-scalar channel) $\sigma(\omega,\bold{p}=\bold{0})$ and thus on the temporal correlation function $G(\tau,\bold{p}=\bold{0})$ obtained through Eq. (\ref{eq:gtau}).\\
Before addressing the issue of improving the HTL results, let us discuss the basic features of the HTL quark spectral function that we want to keep in the new scheme.\\
The structure of the spectral representation given in Eq. (\ref{rappr}) guarantees that:
\beq
\{\,^\star S,\gamma^5\}=0\;,\label{eq:anticomm}
\eeq  
$^\star S$ being the HTL resummed quark propagator. This expresses the fact that, in spite of the appearance in the spectrum of a \emph{gap mass} $m_q$ and of an \emph{asymptotic mass} $\hat{m}_\infty$ for the normal quark mode, both temperature dependent, nevertheless the chirality of a quark propagating in the thermal bath is conserved.\\
One would like Eq. (\ref{eq:anticomm}) to hold even when a better description for the hard fermionic modes is given. Thus, in analogy with Eqs. (\ref{spectral}) and (\ref{fermspec}), the following spectral representation for the NLA quark propagator
\beq
S^{\trm{NLA}}(i\omega_{n},\bold{k})=-\int_{-\infty}^{+\infty}\!d\omega\frac{\rho^{\trm{NLA}}(\omega,\bold{k})}{i\omega_n-\omega}\;,
\eeq
where
\beq
\rho^{\trm{NLA}}(\omega,\bold{k})=\frac{\gamma^0-\bold{\gamma\cdot\hat{k}}}{2}\rho_{+}^{\trm{NLA}}(\omega,k)\, +\, \frac{\gamma^0+\bold{\gamma\cdot\hat{k}}}{2}\rho_{-}^{\trm{NLA}}(\omega,k)\;,\label{eq:specNLA}
\eeq
is assumed to hold. The functions $\rho_\pm^{\trm{NLA}}$ are discussed below.\\
Now the HTL quark spectral functions $\rho_{\pm}$ obey the following \emph{sum rule} \cite{lb}
\beq
\int_{-\infty}^{+\infty}d\omega\,\rho_{\pm}(\omega,k)=1\;,\label{sumprima}
\eeq
which can be cast in the form
\beq
Z_{+}(k)+Z_{-}(k)+\int_{-k}^{+k}d\omega\,\rho_{\pm}(\omega,k)=1\label{quarksomma}\;,
\eeq
where the functions
\beq
Z_{\pm}(k)=\frac{\omega_{\pm}^2(k)-k^2}{2m_q^2}\;
\eeq
are the residues of the quasi-particle excitations.\\
Since Eq. (\ref{sumprima}) is derived directly from the equal-time anticommutation relation:
\beq
\{\psi_\alpha(t,\bold{x}),\psi_\beta^{\dagger}(t,\bold{x'})\}=\delta_{\alpha\beta}\delta(\bold{x}-\bold{x'})\;
\eeq
one would like it to hold for the NLA quark spectral function as well.\\ 
It is of interest to discuss how the saturation of the sum rule (\ref{sumprima}) is implemented in the HTL approximation. To this aim we recall the asymptotic behaviour of the HTL fermionic dispersion relations for $k\gg m_q$ which is found to be
\beq\label{grandi}
\omega_{+}(k)\simeq\sqrt{k^2+\hat{m}_{\infty}^2}\;,\qquad \omega_{-}(k)\simeq k+2k\,\exp\left(-\frac{2k^2+m_q^2}{m_q^2}\right)\;,
\eeq
where the HTL quark asymptotic mass has been defined in Eq. (\ref{eq:htlinf}). Concerning the residues one gets, always for $k\gg m_q$,
\beq
Z_{+}(k)\simeq 1+\frac{m_q^2}{2k^2}\left(1-\ln\frac{2k^2}{m_q^2}\right)\label{resnorm}\;,\quad Z_{-}(k)\simeq\frac{2k^2}{m_q^2}\,\exp\left(-\frac{2k^2+m_q^2}{m_q^2}\right)\;.
\eeq
From the above it follows that:
\beq
\lim_{k\to\infty}Z_+(k)=1\quad\trm{and}\quad \lim_{k\to\infty}Z_-(k)=0\;. 
\eeq
Hence, for $k\gg m_q$ (admittedly a situation where the HTL approximation is no longer realistic) the normal quark mode tends to saturate the sum rule (\ref{quarksomma}) and its dispersion relation approaches the one of a free particle of mass $\hat{m}_\infty$.\\

In line with the above arguments we propose an improved expression for the quark spectral function which satisfies the following constraints:
\begin{itemize}
\item for small momenta it reduces to the HTL one;
\item for large momenta it yields a spectrum dominated by an undamped excitation whose dispersion relation approaches $\epsilon_k^{\trm{NLA}}=\sqrt{k^2+m_\infty^2}$, $m_\infty$ being the solution of Eq. (\ref{massanla});
\item it obeys the sum rule (\ref{sumprima});
\item the associated quark propagator anti-commutes with $\gamma^5$. 
\end{itemize} 
A prescription should also be given to allow a transition from the soft to the hard momenta regime. In this connection, keeping for the intermediate cutoff $\Lambda$ the value given in Eq. (\ref{spatialcutoff}), a natural choice might appear the following one:
\beq
\rho_\pm^{\trm{NLA}}(\omega,k)=\theta(\Lambda-k)\rho_\pm(\omega,k)+\theta(k-\Lambda)\delta(\omega\mp\epsilon_k^{\trm{NLA}})\;,
\eeq
where $\rho_\pm$ are the HTL quark spectral functions, while $\epsilon_k^{\trm{NLA}}=\sqrt{k^2+m_\infty^2}$, being $m_\infty$ the NLA quark asymptotic mass.\\
Actually, since $\omega_+(\Lambda)>\epsilon_k^{\trm{NLA}}(k\!=\!\Lambda)$, a narrow interval of values of $\omega$ (i.e. $2\epsilon^{\trm{NLA}}(\Lambda)<\!\omega\!<2\omega_+(\Lambda)$) exists in which the meson spectral function would get contribution both from the normal quark mode $\omega_+(k)$, namely
\beq
\frac{2N_c}{\pi^2}(e^{\beta\omega}-1)\int_0^\Lambda dk\,k^2\,\tilde{n}^2(\omega_+(k))Z_+^2(k)\delta(\omega-2\omega_+(k))\;,
\eeq
and from the asymptotic quark dispersion relation $\epsilon_k^{\trm{NLA}}$, namely
\beq
\frac{2N_c}{\pi^2}(e^{\beta\omega}-1)\int_\Lambda^\infty dk\,k^2\,\tilde{n}^2(\epsilon_k^{\trm{NLA}})\delta(\omega-2\epsilon_k^{\trm{NLA}})\;.
\eeq
Hence, due to this overcounting of the same physical mode, the meson spectral function would display an unphysical bump for $2\epsilon^{\trm{NLA}}(\Lambda)<\!\omega\!<2\omega_+(\Lambda)$, $\sigma(\omega)$ being discontinuous at the border of the interval.\\
In order to cure this problem (at least partially, as we will see in the following) and to interpolate smoothly between the soft and hard regimes we introduce two additional momenta $\Lambda_1$ and $\Lambda_2$ defined as follows:
\bseq
\begin{align}
\omega_+(\Lambda_1)&=\sqrt{\Lambda^2+m_\infty^2}\\
\omega_+(\Lambda)&=\sqrt{\Lambda_2^2+m_\infty^2}\;.
\end{align}
\eseq
Hence the following guess on the spectral functions $\rho_\pm^{\trm{NLA}}(\omega,k)$ entering into Eq. (\ref{eq:specNLA}), namely
\beq
\rho_\pm^{\trm{NLA}}(\omega,k)=\left\{ \begin{array}{ll}
\rho_\pm(\omega,k) & \trm{if $k<\Lambda_1$}\\
\cos^2(\alpha(k))\rho_\pm(\omega,k)+\sin^2(\alpha(k))\delta(\omega\mp\epsilon_k^{\trm{NLA}}) & \trm{if $\Lambda_1<k<\Lambda_2$}\\
\delta(\omega\mp\epsilon_k^{\trm{NLA}}) & \trm{if $k>\Lambda_2$}
\end{array} \right.\label{eq:ansatzNLA}
\eeq
with
\beq
\alpha(k)=\frac{\pi}{2}\cdot\frac{k-\Lambda_1}{\Lambda_2-\Lambda_1}\label{eq:alphak}\;,
\eeq
appears suitable.\\
Replacing then the HTL quark spectral functions entering into Eq. (\ref{pseudo}) with the above given ansatz, one gets the following NLA expression for the pseudo-scalar meson spectral function (for $\omega>0$):
\begin{multline}
\sigma_{\trm{NLA}}^{\trm{ps}}(\omega,\bold{0})=\frac{2N_c}{\pi^2}(e^{\beta\omega}-1)\!\int_0^{+\infty}dk\, k^2\int_{-\infty}^{+\infty}d\omega_1\int_{-\infty}^{+\infty}d\omega_2 \tilde{n}(\omega_1)\tilde{n}(\omega_2)\\
\delta(\omega\!-\!\omega_1\!-\!\omega_2)[\rho_{+}^{\trm{NLA}}(\omega_1,k)\rho_{+}^{\trm{NLA}}(\omega_2,k)+\rho_{-}^{\trm{NLA}}(\omega_1,k)\rho_{-}^{\trm{NLA}}(\omega_2,k)]\;.\label{pseudoNLA}
\end{multline} 
In conformity with Eq. (\ref{eq:gtau}), one can next obtain the associated zero momentum temporal correlator $G_{\trm{NLA}}^{\trm{ps}}(\tau,\bold{0})$ whose behaviour will be later shown for different values of the temperature.\\ 

\begin{figure}[tp]
\begin{center}
\includegraphics[clip,width=0.6\textwidth]{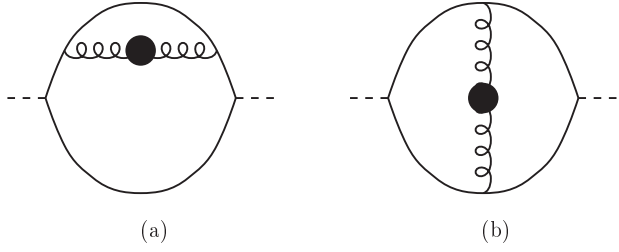}
\caption{Higher order diagrams contributing to the pseudoscalar meson spectral function. All the fermionic lines are hard. The dressed wavy lines are HTL resummed gluon propagators. The effect of diagram (a) (together with the corresponding one with the self energy insertion in the other fermionic line) are taken into account in an effective way through the introduction of the NLA quark effective mass. The effects of diagram (b), which in principle might give a contribution of the same order of (a), are not accounted for in our NLA scheme.}\label{fig:NLA} 
\end{center}
\end{figure}

Before presenting our numerical results we stress that our approach is just an effective way of accounting for the contribution to the meson correlation function arising from the hard fermionic modes, for which the HTL approximation no longer holds. Specifically we embody, in analogy with entropy calculations of Refs. \cite{bie,blasu,blathe,blater}, the effects of diagram (a) in Fig. (\ref{fig:NLA}) into the NLA quark asymptotic mass, which turns out to be lower than the HTL one. On the other hand we neglect the corrections arising from diagram (b) which in principle could be equally important.
%
%
\section{Numerical results}\label{num}
\begin{figure}[htp]
\begin{center}
\includegraphics[clip,width=0.7\textwidth]{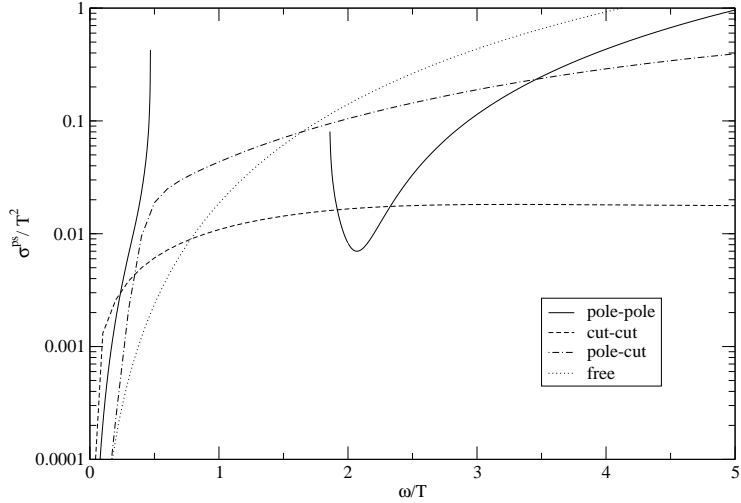}
\caption{The various contributions (pole-pole, pole-cut and cut-cut) to the dimensionless spectral function of a pseudoscalar meson $\sigma^{\trm{ps}}/T^2$ at zero spatial momentum as a function of $x=\omega/T$. Here the quark propagators have been evaluated in the HTL approximation both for hard and soft momenta. In order to test the present results versus the ones obtained in Ref. \cite{kar}, \emph{here} the plot is given for a value of $T$ such that $g(T)=\sqrt{6}$, entailing $m_q=T$. In the pole-pole contribution one recognizes the occurrence of the Van Hove singularities at $x=0.47$ and at $x=1.856$. Also plotted is the free spectral function.}\label{diversi} 
\end{center}
\end{figure}

In this section we first assess the weight of the different contributions (pole-pole, pole-cut and cut-cut) to the HTL pseudoscalar spectral function
\beq
\sigma^{\trm{ps}}_{\trm{HTL}}(\omega,\bold{0})=\sigma^{\trm{pp}}(\omega,\bold{0})+\sigma^{\trm{pc}}(\omega,\bold{0})+\sigma^{\trm{cc}}(\omega,\bold{0})\;.
\eeq
Next we explore the modifications of the HTL results stemming from the more proper NLA treatment of the hard modes.\\
For the sake of consistency, we start from the explicit expression for the three terms involved in $\sigma^{\trm{ps}}_{\trm{HTL}}(\omega,\bold{0})$. These have already been deduced in Ref. \cite{kar}.\\
For zero spatial momentum the pole-pole term, which selects the quasi-particle peaks in both quark spectral functions in Eq. (\ref{pseudo}), is given by:
\begin{multline}
\sigma^{\trm{pp}}(\omega)=\frac{N_c}{2\pi^2}\frac{e^{\beta\omega}-1}{m_q^4}\left[\tilde{n}^2(\omega_{+}(k_1))(\omega_{+}^2(k_1)-k_1^2)^2\frac{k_1^2}{2|\omega'_{+}(k_1)|}+\right.\\
+2\!\sum_{k_2}\tilde{n}(\omega_{+}(k_2))[1-\tilde{n}(\omega_{-}(k_2))](\omega_{+}^2(k_2)-k_2^2)(\omega_{-}^2(k_2)-k_2^2)\frac{k_2^2}{|\omega'_{+}(k_2)\!-\!\omega'_{-}(k_2)|}+\\
\left.+\sum_{k_3}\tilde{n}^2(\omega_{-}(k_3))(\omega_{-}^2(k_3)-k_3^2)^2\frac{k_3^2}{2|\omega'_{-}(k_3)|}\right]\label{polepole}\;.
\end{multline}
In the above $k_1$ is the unique solution of the equation $\omega-2\omega_{+}(k_1)=0$, whereas two are the values of $k_2$ fulfilling $\omega-\omega_{+}(k_2)+\omega_{-}(k_2)=0$ and, likewise, two values of $k_3$ satisfy $\omega-2\omega_{-}(k_3)=0$.\\
In order to identify the different contributions to the pole-pole term with well defined physical processes let us introduce the notation $q_+$ ($\bar{q}_+$) for the normal (anti-) quark mode, $q_-$ ($\bar{q}_-$) for the (anti-) plasmino mode and $M$ for the pseudo-scalar meson.\\
Thus the first term in Eq. (\ref{polepole}) corresponds to the annihilation of a normal quark anti-quark pair into a meson at rest
\begin{displaymath}
q_+\;+\;\bar{q}_+\longrightarrow M\;,
\end{displaymath}
which clearly turns out to be proportional to the mean occupation number of normal (anti-) quarks states, hence the factor $\tilde{n}^2(\omega_{+}(k_1))$. Due to energy conservation the density of states turns out to be proportional to $|d(2\omega_+)/dk|^{-1}$.\\
On the other hand the second term is related to the decay of a normal quark mode into a plasmino mode with the same momentum and a meson at rest, namely
\begin{displaymath}
q_+\longrightarrow {q}_- +M\;.
\end{displaymath}
Therefore in this term one of the statistical factor accounts for the mean occupation number of normal quarks states (hence the Fermi distribution $\tilde{n}(\omega_{+}(k_2))$), while the other one expresses the fact that the final plasmino has to occupy empty states (hence the Pauli-blocking factor $1-\tilde{n}(\omega_{-}(k_2))$). In this case the density of states is proportional to $|d(\omega_+-\omega_-)/dk|^{-1}$.\\
Finally the third term describes the annihilation of a plasmino anti-plasmino pair into a meson
\begin{displaymath}
q_-\;+\;\bar{q}_-\longrightarrow M\;,
\end{displaymath}
whence the statistical $\tilde{n}^2(\omega_{-}(k_3))$ and the density of states $|d(2\omega_-)/dk|^{-1}$ factors.\\
It is important to realize that, owing to the minimum in the plasmino dispersion relation, two values of $\omega$ exist for which the denominators in Eq. (\ref{polepole}) $|\omega'_{+}(k_2)\!-\!\omega'_{-}(k_2)|$ and $|\omega'_{-}(k_3)|$ vanish, thus yielding singularities in the meson spectral function . These are the so-called Van Hove singularities, essentially reflecting a divergence in the density of states. One finds (numerically) that they occur at $\omega\approx 0.47m_q$ and at $\omega\approx 1.86m_q$. The impact of these singularities on the meson spectral function is dramatic for soft values of $\omega$. Here indeed the HTL meson spectral function turns out to be orders of magnitude larger than the free one, as can be seen in Fig. (\ref{diversi}).\\
Turning to the pole-cut contribution (for positive energies) it is found to be:
{\setlength\arraycolsep{1pt}
\beqa
\sigma^{\trm{pc}}(\omega) & = & \frac{2N_c}{\pi^2}\frac{e^{\beta\omega}-1}{m_q^2}\int_{0}^{\infty}dk\,k^2\,\cdot\nonumber\\
{} & {} & \cdot\left[\theta(k^2-(\omega-\omega_{+})^2)\tilde{n}(\omega-\omega_+)\tilde{n}(\omega_+)\beta_{+}(\omega-\omega_+,k)(\omega_+^2-k^2)\right.\nonumber\\
{} & {} & \left.+\theta(k^2-(\omega-\omega_{-})^2)\tilde{n}(\omega-\omega_-)\tilde{n}(\omega_-)\beta_{-}(\omega-\omega_-,k)(\omega_-^2-k^2)\right]\,.\nonumber\\
\eeqa}
Finally the cut-cut contribution is given by:
\begin{multline}
\sigma^{\trm{cc}}(\omega)=\frac{2N_c}{\pi^2}(e^{\beta\omega}-1)\int_0^\infty dk\,k^2\int_{-k}^{+k}dx\,\tilde{n}(x)\tilde{n}(\omega-x)\theta(k^2-(\omega-x)^2)\cdot\\
\cdot\left[\beta_+(x,k)\beta_+(\omega-x,k)+\beta_-(x,k)\beta_-(\omega-x,k)\right]\,.
\end{multline}
The three above quoted contributions to the meson spectral function in the pseudoscalar channel in a pure HTL approximation are plotted in Fig. \ref{diversi}: they numerically coincide with the results previously obtained by Karsch et al. \cite{kar}.\\     

The question we now raise is whether the behaviour of the meson spectral function at soft energies, displayed in Fig. \ref{diversi}, is qualitatively modified once a better description of the hard fermionic modes is adopted.\\
Since the divergence in the density of states, which gives rise to the Van Hove singularities in the pole-pole term, stems from the stationary points of the curves given by the equations 
\bseq
\begin{align}
\omega&=2\omega_{-}(k)\label{prima}\\
\trm{and}&{}\nonumber\\
\omega&=\omega_+(k)-\omega_{-}(k)\label{seconda}\;,
\end{align}
\eseq
to answer the above question one should check whether the choice of the intermediate cutoff $\Lambda$ separating the soft modes from the hard ones yields $\Lambda>\tilde{k}_i$, where $\tilde{k}_i$ are the stationary points of the curves (\ref{prima}) and (\ref{seconda}). In such a case indeed the Van Hove singularities will still be present in the spectral function. Therefore we report below the numerical values of $\Lambda_1,\Lambda,\Lambda_2$ (normalized to the quark thermal mass $m_q$) for two different choices of the intermediate cutoff $\Lambda$, for $T=2T_c$ and $T=4T_c$. We also report the corresponding value of the gauge coupling $g$ obtained through the QCD $\beta$-function as discussed below.\\

\begin{center}
\begin{tabular}{|c|c|c|c|}
\hline
$T=2T_c$ ($g$=1.785) & $\Lambda_1/m_q$ & $\Lambda/m_q$ & $\Lambda_2/m_q$\\
\hline
$\Lambda=\sqrt{\pi T\hat{m}_d}$ & 3.312 & 3.492 & 3.665\\
\hline
$\Lambda=\sqrt{2\pi T\hat{m}_d}$ & 4.806 & 4.939 & 5.069\\
\hline
\end{tabular}
\end{center}

\begin{center}
\begin{tabular}{|c|c|c|c|}
\hline
$T=4T_c$ ($g$=1.548) & $\Lambda_1/m_q$ & $\Lambda/m_q$ & $\Lambda_2/m_q$\\
\hline
$\Lambda=\sqrt{\pi T\hat{m}_d}$ & 3.593 & 3.750 & 3.902\\
\hline
$\Lambda=\sqrt{2\pi T\hat{m}_d}$ & 5.188 & 5.303 & 5.417\\
\hline
\end{tabular}
\end{center}
 
\begin{figure}[htp]
\begin{center}
\includegraphics[clip,width=0.7\textwidth]{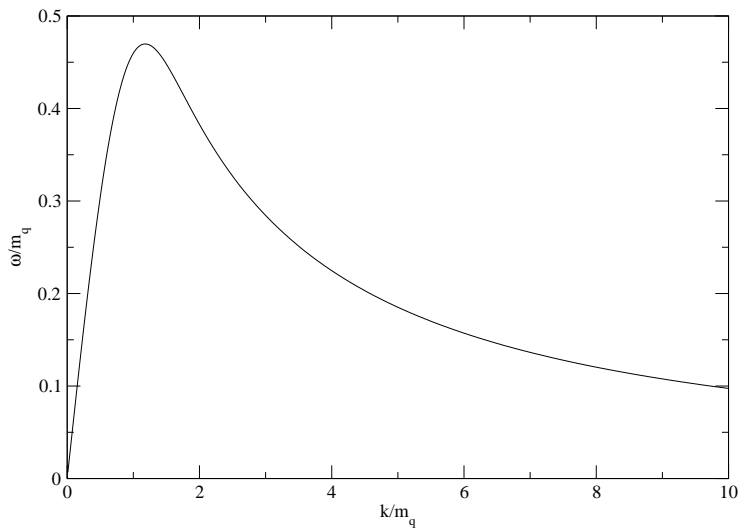}
\caption{The behaviour of the difference between the two fermionic dispersion relations $\tilde{\omega}_{+}(x)-\tilde{\omega}_{-}(x)$ . The dimensionless variables $\tilde{\omega}=\omega/m_q$ and $x=k/m_q$ are used. The maximum of the curve occurs at $x=1.179$ and corresponds to a value of $\tilde{\omega}=0.4698$.}\label{diffe}
\end{center}
\end{figure}

From the comparison with the curves in Figs. \ref{fermdisp} and \ref{diffe} which, being normalized to the quark thermal mass $m_q$, hold for every value of the temperature, one realizes that the stationary points leading to the Van Hove singularities \textit{always} occur in the soft domain $k<\Lambda_1$. Hence we conclude that, \textit{qualitatively}, the behaviour of the HTL meson spectral function does not change once a better treatment of the hard fermionic modes is kept into account. The quantitative changes are below examined.\\

Accordingly, we next present the results obtained in the Next to Leading Approximation (NLA), which has been briefly discussed in the previous section and which allows a better treatment of the contribution of the hard quark modes in the meson spectral function.\\ 

Before addressing the discussion of the numerical results a few words, regarding how the running of $\alpha_s$ is related to the temperature, are in order.\\
The running of the QCD coupling $g$ with the renormalization scale $\bar{\mu}$ (introduced in the dimensional regularization scheme) is here obtained through the two-loop $\beta$-function with $N_f=2$, in accord with \cite{bie}. The modified Minimal Subtraction ($\overline{\trm{MS}}$) renormalization scheme is employed. Dealing with massless quarks, the only physical scale entering into the problem is the temperature. Hence the evaluation of the coupling $g$ at the scale $\bar{\mu}=2\pi T$, which corresponds to the spacing between the Matsubara frequencies, appears a reasonable choice. For further details see \cite{bie} and references therein.\\ 

\begin{figure}[htp]
\begin{center}
\includegraphics[clip,width=0.7\textwidth]{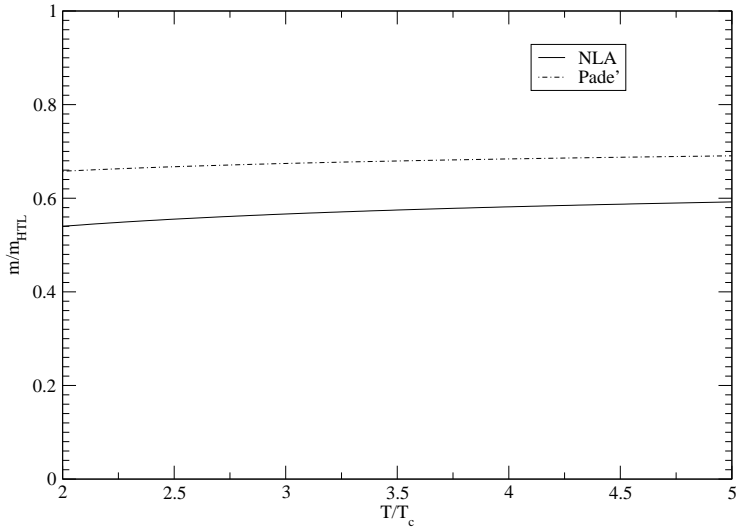}
\caption{Ratio of the asymptotic quark mass over its HTL value as a function of $T/T_c$: here the (negative) correction $\delta m_{\infty}^2$ has been included in two different ways: with the first Pade' approximant (dot-dashed line) and in the ``self-consistent like'' way (continuous line).}\label{ratio}   
\end{center}
\end{figure}

\begin{figure}[htp]
\begin{center}
\includegraphics[clip,width=0.7\textwidth]{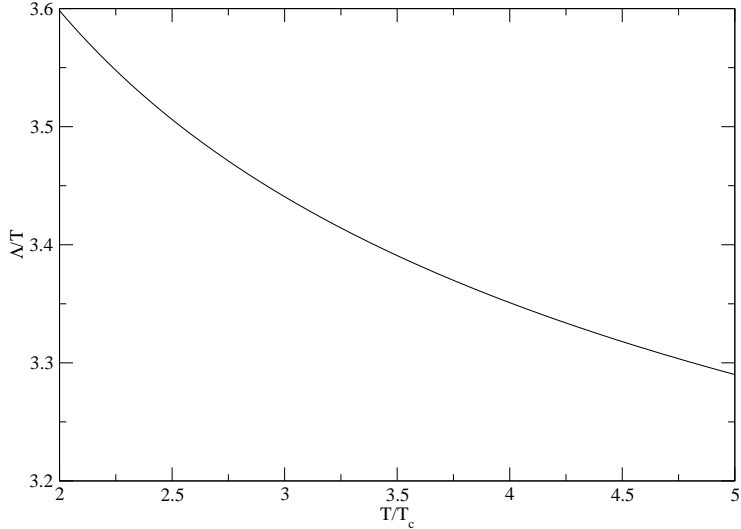}
\caption{Temperature behaviour of the intermediate cutoff $\Lambda=\sqrt{2\pi T\hat{m}_D}$ separating soft and hard momenta. Dimensionless units $\Lambda/T$ and $T/T_c$ are adopted. The decreasing of the ratio $\Lambda/T$ reflects the running of the strong coupling constant.}\label{cutoff}
\end{center}
\end{figure}

In Fig. \ref{ratio} we display the ratio of the Pade' and NLA asymptotic mass with respect to its HTL value. The two curves represent the solution of Eqs. (\ref{massapade'}) and (\ref{massanla}). It appears clearly that the interaction of the hard quark modes with soft longitudinal and transverse gluons (for which one uses HTL resummed propagators) lowers the value of the asymptotic quark mass and that the two approximation schemes lead to similar results.\\
In Fig. \ref{cutoff} the behaviour of the ratio $\Lambda/T$ as a function of the temperature is given.\\
  
\begin{figure}[htp]
\begin{center}
\includegraphics[clip,width=1.0\textwidth]{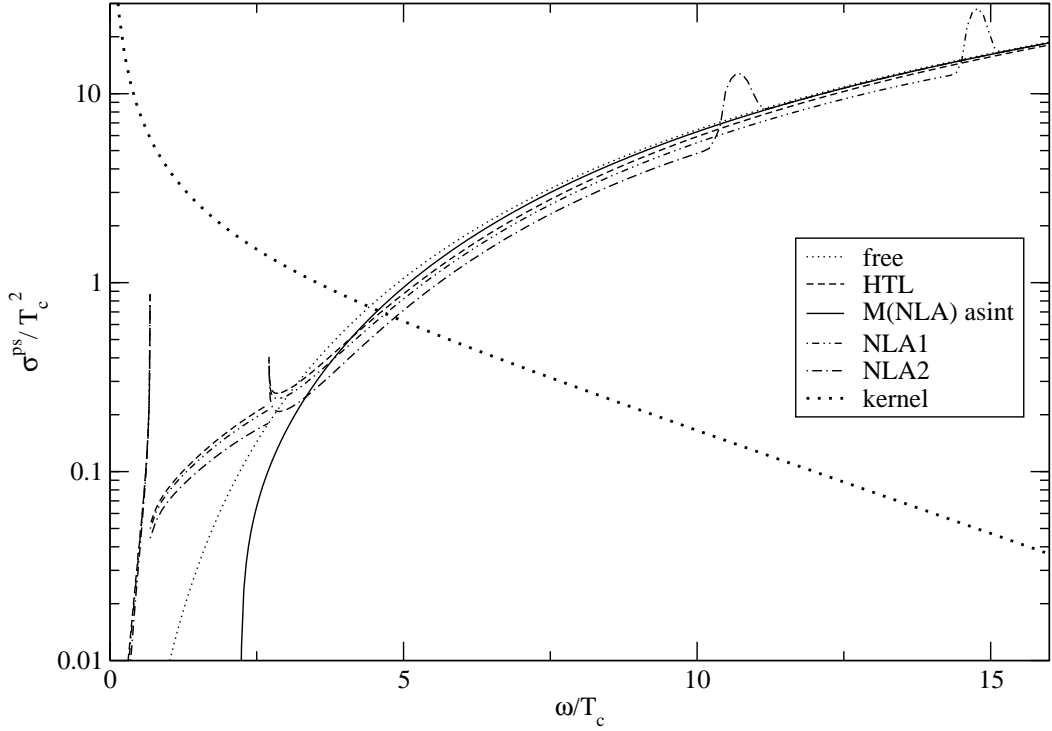}
\caption{Behaviour of the zero momentum pseudoscalar spectral function $\sigma^{ps}/T_c^2$ versus $\omega/T_c$ in different approximations: free result, HTL, NLA and quarks endowed with a thermal mass $m_\trm{NLA}$. NLA1 corresponds to the choice $\Lambda=\sqrt{2\pi T \hat{m}_D}$, NLA2 to $\Lambda=\sqrt{\pi T \hat{m}_D}$. The plot refers to $T=2T_c$.}\label{sommanla}
\end{center}
\end{figure}

In Fig. \ref{sommanla} we plot the pseudoscalar meson spectral function obtained in the NLA scheme with two different choices of the cut-off parameter $\Lambda$. This is compared with the pure HTL result, with the curve obtained assigning to the quarks a mass $m_\infty^{\trm{NLA}}$ and with the non interacting case. Notice that the curves NLA1 and NLA2 display a peak which has no physical meaning, but is an artifact of our approximation for the quark spectral function. In fact a narrow energy interval exists where the meson spectral function gets contribution both from the normal HTL quark mode and from the NLA asymptotic dispersion relation. On the other hand, as discussed in the previous section, without the prescription (\ref{eq:ansatzNLA}) the spectral function would display an even worse unphysical behaviour. Although our findings for this intermediate scale of frequencies cannot be trusted, nevertheless for the soft and hard regimes they are reliable.\\
In Fig. \ref{sommanla} the kernel entering into the definition of the meson temporal correlation function (see formula (\ref{eq:gtau})), namely
\beq
K(\tau,\omega)=\cosh(\omega(\tau-\beta/2))/\sinh(\omega\beta/2)\;,
\eeq
is also plotted for $\tau=\beta/2$. One infers from the behaviour of $K(\tau=\beta/2,\omega)$ that, for this value of $\tau$, the contribution from the soft sector to the integral in Eq. (\ref{eq:gtau}) is significant.\\

\begin{figure}[htp]
\begin{center}
\includegraphics[clip,width=1.0\textwidth]{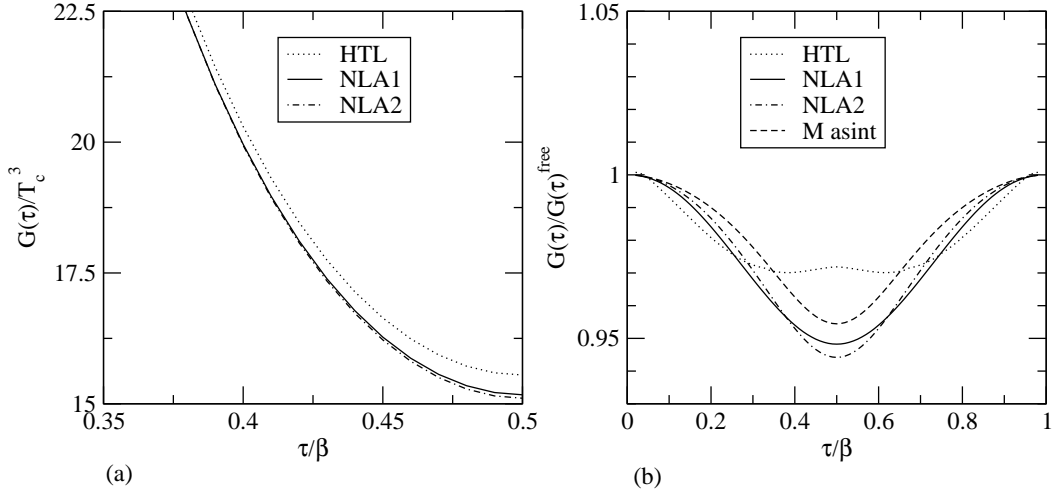}
\caption{(a): Behaviour of $G(\tau)/T_c^3$ vs $\tau/\beta$. (b): Behaviour of $G(\tau)/G^{\trm{free}}(\tau)$ vs $\tau/\beta$. NLA1 corresponds to $\Lambda=\sqrt{2\pi T \hat{m}_D}$, NLA2 corresponds to $\Lambda=\sqrt{\pi T \hat{m}_D}$. In panel (b) we also display the result obtained in the case of quarks endowed with a thermal mass $m=m_\trm{NLA}$. The curves are given for $T=2T_c$.}\label{doublepanel2}
\end{center}
\end{figure}

\begin{figure}[htp]
\begin{center}
\includegraphics[clip,width=1.0\textwidth]{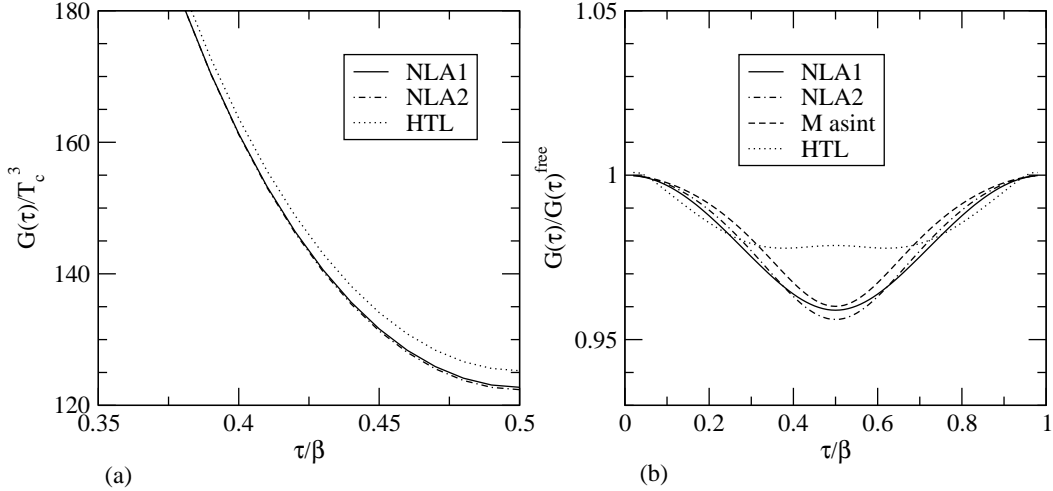}
\caption{The same as in Fig. \ref{doublepanel2}, but for $T=4T_c$.}\label{doublepanel4}
\end{center}
\end{figure}

Finally in Figs. \ref{doublepanel2} and \ref{doublepanel4} we show, for two different values of the temperature, the zero-momentum temporal correlator $G(t=-i\tau,\bold{p}=\bold{0})$ in the two different approximations schemes discussed in this paper: HTL and NLA. We also show the curve obtained dressing the quark modes with the NLA asymptotic mass and setting to zero the plasmino and the damping contribution to the quark spectral function.\\
As expected from the general theory, the curves are symmetric for $\tau\,\to\,\beta-\tau$.\\
The correlation function diverges for $\tau\to 0\;(\beta)$ independently from the approximation scheme employed and even in the non-interacting case. Hence the meaningful quantity to study in order to assess the impact of the interaction with the thermal bath on the meson correlation function is the ratio $G(\tau)/G^{\trm{free}}(\tau)$ given in panel B in Figs. \ref{doublepanel2} and \ref{doublepanel4}. Note that for $\tau\to 0(\beta)$ one recovers the free result.\\
For $\tau$ sufficiently small (or close to $\beta$) the NLA curves stay above the HTL result. This may be understood on the following basis. As $\tau$ gets smaller, the contribution to $G(\tau)$ coming from large frequencies grows, as one can see from Eq. (\ref{eq:gtau}). But for large values of $\omega$ the meson spectral function (both in HTL and in NLA) is dominated by the normal quark modes, in which $m_\infty^{\trm{(NLA)}}<\hat{m}_\infty^{\trm{(HTL)}}$.\\
On the contrary, for $\tau$ close to $\beta/2$, where the contribution from soft frequencies becomes more important, the HTL result stays significantly above the NLA curves.\\
It is also evident that the NLA curve are not much sensitive to the choice of the cutoff $\Lambda$ and that, as one moves to larger values of the temperature, one approaches (slowly) the free result.\\
In any case, in the whole range $0<\tau<\beta$, is $G(\tau)<G^{\trm{free}}(\tau)$. This outcome appears to agree with the results for the correlation function along the $z$-axis obtained in \cite{lai}. Here, on the basis of a dimensionally reduced effective lagrangian, a positive correction (of the order of $5\%$) was found to occur for the meson screening mass.  

\section{Conclusions}
In this paper we have explored the predictions of the NLA framework on the thermal meson correlation functions in the deconfined phase of QCD, thus extending past work of Blaizot et al. who successfully investigated, employing the NLA scheme, the thermodynamical properties of the QGP. Notably their results turned out to be well supported by lattice calculations, even in a regime of temperatures where $g$ is not small ($g\sim 1$ or larger), in which the separation of different momentum scales (i.e. $g^2T,\; gT, \; T$) might appear questionable. Hence we felt justified to employ here the same scheme to deal with other observables, like the meson correlation functions.\\
Actually what we have proposed is a variant of the NLA scheme, smoothly matching the soft and hard momenta regimes.\\
With a more careful treatment of the contribution to the meson correlation function arising from the hard quark modes we have still found that the NLA does not change dramatically the HTL results which turns out to be quite robust. In particular the peculiar behaviour of the meson spectral function for soft energies (Van-Hove singularities) remains (almost) unaltered.\\
The detection of the V.H. singularities in the meson spectral functions, being related to the minimum in the plasmino dispersion relation, would be a clear signal of deconfinement. Indeed the doubling of the fermionic dispersion relation for non-vanishing three-momentum is not strictly related to the HTL approximation, whose validity, for values of the temperature just above $T_c$, is questionable. Actually, as discussed in \cite{pesh,wel}, it is possible to show, in a totally non perturbative framework, that in the high temperature chirally symmetric phase of QCD the fermion propagator \textit{does always display} two distinct dispersion relations, one of which developing a minimum.\\    
Unfortunately the possibility of getting an experimental evidence of these singularities, for instance from the dilepton production rate for what concerns the vector channel, is dim, owing to the fact that during the expansion of the fireball their position moves toward lower energies as the temperature decreases (experiments provide time-integrated yields). Furthermore there are other sources of dileptons which may cover the signal.\\
At present, lattice calculations of (zero momentum) meson spectral functions (of light quarks) $\sigma_M(\omega)$ do not show any evidence of sharp resonances in the soft energy domain which could be related to the V.H. singularities \cite{bile1,bile2,bile3}. This occurrence may depend on the fact that 
what one actually measures on the lattice is the temporal correlator $G_M(\tau)$ on a finite set of euclidean times. Hence, to get $\sigma_M(\omega)$ from the lattice requires to invert Eq. (\ref{eq:gtau}) and the left hand side of this equation is known only on a finite set of values of $\tau$. To perform this inversion, the Maximum Entropy Method has been recently adopted \cite{dat,asa1,asa2,bile1,bile2,bile3}, which in any case requires some \textit{a priori} information both on $G_M(\tau)$ and on $\sigma_M(\omega)$. As a consequence it is not possible at present to assure that the spectral function reconstructed in this way is unique and that the existence of very sharp resonances like the V.H. singularities can be brought to evidence with this method.\\
Also of interest is to point out that the V.H. singularities, associated to the vanishing of the group velocity of a propagating wave, also occur in other contests in many-body physics. For example in a non relativistic ferromagnetic system at $T=0$ undergoing a spontaneous symmetry breaking (\emph{Quantum Phase Transition}), the dispersion relation of the Goldstone mode (magnon) displays a maximum, at which the group velocity is zero \cite{ber}.\\   
A promising route to the study of mesonic correlations in the deconfined phase of QCD is offered by the possibility of applying the same strategy developed in this paper to the evaluation of meson screening masses (i.e. considering the propagation along a spatial direction) and comparing the results with those obtained in different approaches (for instance on the lattice or with effective lagrangians). Indeed, due to the larger number of sites along the spatial directions with respect to the (euclidean) temporal one, lattice calculations should provide more reliable results for this kind of observables.\\
Concerning the shortcoming of the framework employed in this paper, it should be pointed out that we took into account corrections to the free meson correlation functions arising merely from dressing the fermionic propagators, due to their interactions with the medium (i.e. the gluons of the thermal bath). On the other hand, if, within the HTL scheme, one is allowed to leave the pseudoscalar vertex untouched (since the HTL correction to the pseudoscalar vertex vanishes), this is no longer true when one tries to go beyond the HTL approximation. While we have not accounted for this fact in our analysis, it certainly represents an item worth to be addressed.\\ 
Equally important appears to be the study of the chiral and of the quark number susceptibilities in a scheme improving upon the HTL results obtained in Refs. \cite{chira,qns}.\\ 
Finally we stress that we don't expect the framework employed in this paper to give reliable results for temperature values lower than $2.5\div 3 T_c$. For the range of $T$ just above $T_c$ other schemes \cite{sh,sh2}, predicting the presence of bound states also in the deconfined phase, appear more realistic. 

\section*{Acknowledgments}
We are grateful to Prof. J.P. Blaizot for encouragements and criticisms on some aspects of the paper. We also thank Dr. A. De Pace and Dr. A. Ipp for fruitful discussions.

\appendix
\section{Spectral representation of the free quark propagator}\label{a}
In configuration space the Matsubara fermionic propagator is defined as follows:
\beqa
S_{\alpha\beta}(\tau,\bold{x};\tau',\bold{x'}) & = & \langle T_\tau(\psi_{\alpha}(-i\tau,\bold{x})\bar{\psi}_{\beta}(-i\tau',\bold{x'}))\rangle\nonumber\\
{} & = &\theta(\tau-\tau')\langle\psi_{\alpha}(-i\tau,\bold{x})\bar{\psi}_{\beta}(-i\tau',\bold{x'})\rangle\nonumber\\
{} & {} & -\theta(\tau'-\tau)\langle\bar{\psi}_{\beta}(-i\tau',\bold{x'})\psi_{\alpha}(-i\tau,\bold{x})\rangle\;,
\eeqa
whereas in Fourier space, for the non interacting Dirac field, it reads:
\beqa
S_{F}(i\omega_n,\bold{p}) & = & -\frac{p\hspace{-.15cm}{\slash}+m}{p^2-m^2}\nonumber\\
{} & = & \frac{p\hspace{-.15cm}{\slash}+m}{\omega_n^2+E_p^2}\label{propferm}\;.\eeqa
In the above $p_0=i\omega_n$, $E_p^2=\bold{p}^2+m^2$ and, as in Blaizot and Iancu in \cite{bi}, the matrices $\gamma^{\mu}$ ($\mu=0,\dots 3$) satisfy the usual Dirac algebra; furthermore we work with a Minkowskian metric. Note that this choice leads to some differences with respect to the expressions obtained with the notations used in \cite{lb}. Of course the quark propagator is diagonal in the flavour and colour indices and the antiperiodic boundary conditions entail:
\beq
S_{\alpha\beta}(\tau-\beta,\bold{x};\tau',\bold{x'})=-S_{\alpha\beta}(\tau,\bold{x};\tau',\bold{x'})\;.
\eeq
Thus the Matsubara frequencies are given by $\omega_n=(2n+1)\pi T$.\\

In terms of the free fermionic spectral function
\beq
\rho_F(p)=\epsilon(p_0)(p\hspace{-.15cm}{\slash}+m)\delta(p^2-m^2)\;,
\eeq
the following representation holds for the free thermal propagator:
\beq\label{freespec}
S_{F}(i\omega_n,\bold{p})=-\int_{-\infty}^{+\infty}dp_0\frac{\rho_F(p_0,\bold{p})}{i\omega_n-p_0} \qquad .
\eeq
It may be useful, in some computations, to use the following expression for the fermionic propagator in the mixed representation:
\beq\label{mixed}
S_F(\tau\!>0,\bold{p})=\int_{-\infty}^{+\infty}dp_0\,e^{-p_0\tau}\rho_F(p_0,\bold{p})(1-\tilde{n}(p_0))\;.
\eeq
It is easily verified that Eq. (\ref{mixed}) leads to the proper spectral representation (\ref{freespec}). In fact:
\begin{multline}
S_F(i\omega_n,\bold{p})=\!\int_{0}^{\beta}\!\!d\tau e^{i\omega_n\tau}S_F(\tau,\bold{p})=\!\int_{0}^{\beta}\!\!d\tau e^{i\omega_n\tau}\!\int_{-\infty}^{+\infty}\!\!dp_0\,e^{-p_0\tau}\rho_F(p)(1\!-\!\tilde{n}(p_0))\\
=\int_{-\infty}^{+\infty}\!\!dp_0\rho_F(p_0,\bold{p})(1\!-\!\tilde{n}(p_0))\int_{0}^{\beta} d\tau e^{(i\omega_n-p_0)\tau}=-\int_{-\infty}^{+\infty}dp_0\frac{\rho_F(p_0,\bold{p})}{i\omega_n-p_0}\,.
\end{multline} 

\section{HTL resummed quark propagator}\label{b}
In this appendix, following \cite{bi}, we quote the spectral representation for the HTL resummed quark propagator which is suitable for performing loop calculations and also helpful in understanding the physical meaning of the different terms one gets in the final result.\\
In the massless case the quark propagator in HTL approximation can be written as follows \cite{bi}:
\beq\label{delta}
^{\star}S(\omega,\bold{p})=\,^{\star}\Delta_{+}(\omega,p)\frac{\gamma^0-\bold{\gamma\cdot\hat{p}}}{2}+\,^{\star}\Delta_{-}(\omega,p)\frac{\gamma^0+\bold{\gamma\cdot\hat{p}}}{2}
\eeq
being
\beq\label{den}
^{\star}\Delta_{\pm}(\omega,p)=\frac{-1}{{\displaystyle\omega\mp p-\frac{m_q^2}{2p}\left[\left(1\mp \frac{\omega}{p}\right)\ln\frac{\omega+p}{\omega-p}\pm 2\right]}}\; ,
\eeq
with the quark thermal mass $m_q(T)\!=\!(gT)/\sqrt{6}$. Likewise the following representation
\beq\label{rappr}
^{\star}S(i\omega_n,\bold{p})=-\frac{\gamma^0\!-\!\bold{\gamma\cdot\hat{p}}}{2}\!\int_{-\infty}^{+\infty}\!d\omega\frac{\rho_{+}(\omega,p)}{i\omega_n-\omega}\, -\, \frac{\gamma^0\!+\!\bold{\gamma\cdot\hat{p}}}{2}\!\int_{-\infty}^{+\infty}\!d\omega\frac{\rho_{-}(\omega,p)}{i\omega_n-\omega}\; ,
\eeq
is easily proved to hold as well. In (\ref{rappr}) $\rho_{\pm}$ and $^{\star}\Delta_{\pm}$ are related as follows:
\beq\label{defspectral}
^{\star}\Delta_{\pm}(z,p)=-\int_{-\infty}^{+\infty}\!d\omega\frac{\rho_{\pm}(\omega,p)}{z-\omega}\, \Rightarrow \rho_{\pm}(\omega,p)=\frac{1}{\pi}\trm{Im}\, ^{\star}\Delta_{\pm}(\omega+i\eta,p)\; ,
\eeq
$z$ spanning the whole complex plane.\\
Lumping together the above results one gets the following compact expression for the HTL quark spectral function
\beq\label{spectral}
\rho_{\trm{HTL}}(\omega,\bold{p})=\frac{\gamma^0-\bold{\gamma\cdot\hat{p}}}{2}\rho_{+}(\omega,p)\, +\, \frac{\gamma^0+\bold{\gamma\cdot\hat{p}}}{2}\rho_{-}(\omega,p) \; ,
\eeq
with the corresponding spectral representation:
\beq\label{fermspec}
^{\star}S(i\omega_{n},\bold{p})=-\int_{-\infty}^{+\infty}\!d\omega\frac{\displaystyle{\rho_{\trm{HTL}}(\omega,\bold{p})}}{i\omega_n-\omega}\;,
\eeq
for the thermal fermionic Matsubara propagator. This is conveniently used in performing loop calculations. 

\section{Evaluation of $G_{\trm{NLA}}^{\trm{ps}}(\tau,\bold{p}=\bold{0})$}
Here we explicitly write the NLA pseudoscalar meson propagator at zero spatial momentum $G_{\trm{NLA}}^{\trm{ps}}(\tau,\bold{p}=\bold{0})$ split into the different contributions which should be numerically evaluated. Recalling Eq. (\ref{eq:ansatzNLA}) we divide the integration over $k$ into three regions corresponding to soft (a), intermediate (b) and hard (c) momenta.
Starting from the soft momenta sector we have (as usual) three contributions of different kind: pole-pole, pole-cut and cut-cut. Their expressions are (writing $\omega_\pm$ for $\omega_\pm(k)$ and $\tilde{\tau}$ for $\tau/\beta$):
\beqa
G_{(a)}^{\trm{pp}}(\tau) & = & \frac{2N_c}{\pi^2}\int_{0}^{\Lambda_1}k^2\,dk\left\{\left(\frac{1}{1+e^{-\beta\omega_+}}\right)^2\cdot\left(\frac{\omega_+^2-k^2}{2m_q^2}\right)^2\cdot\right.\nonumber\\
{} & {} & \cdot\left[\displaystyle{e^{\displaystyle{-2\beta\omega_+\tilde{\tau}}}}+\displaystyle{e^{\displaystyle{-2\beta\omega_+(1-\tilde{\tau})}}}\right]+\nonumber\\
{} & {} & +2\cdot\left(\frac{1}{1+e^{-\beta\omega_+}}\right)\cdot\left(\frac{1}{1+e^{+\beta\omega_-}}\right)\cdot\left(\frac{\omega_+^2-k^2}{2m_q^2}\right)\cdot\left(\frac{\omega_-^2-k^2}{2m_q^2}\right)\cdot\nonumber\\
{} & {} & \left[\displaystyle{e^{\displaystyle{-\beta(\omega_+-\omega_-)\tilde{\tau}}}}+\displaystyle{e^{\displaystyle{-\beta(\omega_+-\omega_-)(1-\tilde{\tau})}}}\right]+\nonumber\\
{} & {} & +\left.\left(\frac{1}{1+e^{-\beta\omega_-}}\right)^2\left(\frac{\omega_-^2-k^2}{2m_q^2}\right)^2\left[\displaystyle{e^{\displaystyle{-2\beta\omega_-\tilde{\tau}}}}+\displaystyle{e^{\displaystyle{-2\beta\omega_-(1-\tilde{\tau})}}}\right]\right\}\;,\nonumber\\
\label{eq:pphtl}
\eeqa
\beqa
G_{(a)}^{\trm{pc}}(\tau) & = & \frac{2N_c}{\pi^2}\int_{0}^{\Lambda_1}k^2\,dk\int_{-k}^{k}d\omega_2\left\{\frac{1}{1+e^{-\beta\omega_+}}\cdot\frac{1}{1+e^{-\beta\omega_2}}\cdot\right.\nonumber\\
{} & {} & \cdot\frac{\omega_+^2-k^2}{m_q^2}\beta_+(\omega_2,k)\left[\displaystyle{e^{\displaystyle{-\beta(\omega_++\omega_2)\tilde{\tau}}}}+\displaystyle{e^{\displaystyle{-\beta(\omega_++\omega_2)(1-\tilde{\tau})}}}\right]+\nonumber\\
{} & {} & +\frac{1}{1+e^{-\beta\omega_-}}\cdot\frac{1}{1+e^{-\beta\omega_2}}\cdot\frac{\omega_-^2-k^2}{m_q^2}\cdot\beta_-(\omega_2,k)\cdot\nonumber\\
{} & {} & \left.\cdot\left[\displaystyle{e^{\displaystyle{-\beta(\omega_-+\omega_2)\tilde{\tau}}}}+\displaystyle{e^{\displaystyle{-\beta(\omega_-+\omega_2)(1-\tilde{\tau})}}}\right]\right\}
\label{eq:pchtl}
\eeqa
and
\beqa
G_{(a)}^{\trm{cc}}(\tau) & = & \frac{2N_c}{\pi^2}\int_{0}^{\Lambda_1}k^2\,dk\int_{-k}^{k}d\omega_1\int_{-\omega_1}^k d\omega_2\frac{1}{1+e^{-\beta\omega_1}}\cdot\frac{1}{1+e^{-\beta\omega_2}}\cdot\nonumber\\
{} & {} & \cdot[\beta_+(\omega_1,k)\beta_+(\omega_2,k)+\beta_-(\omega_1,k)\beta_-(\omega_2,k)]\cdot\nonumber\\
{} & {} & \cdot\left[\displaystyle{e^{\displaystyle{-\beta(\omega_1+\omega_2)\tilde{\tau}}}}+\displaystyle{e^{\displaystyle{-\beta(\omega_1+\omega_2)(1-\tilde{\tau})}}}\right]\label{eq:cchtl}\;.
\eeqa
The intermediate momentum sector is more involved. We have:
\beqa
G_{(b)}^{\trm{pp}}(\tau) & = & \frac{2N_c}{\pi^2}\int_{\Lambda_1}^{\Lambda_2}k^2\,dk\left\{\left(\frac{1}{1+e^{-\beta\omega_+}}\right)^2\cdot\left(\frac{\omega_+^2-k^2}{2m_q^2}\right)^2\cdot\right.\nonumber\\
{} & {} & \cdot\left[\displaystyle{e^{\displaystyle{-2\beta\omega_+\tilde{\tau}}}}+\displaystyle{e^{\displaystyle{-2\beta\omega_+(1-\tilde{\tau})}}}\right]+\nonumber\\
{} & {} & +2\cdot\left(\frac{1}{1+e^{-\beta\omega_+}}\right)\cdot\left(\frac{1}{1+e^{+\beta\omega_-}}\right)\cdot\left(\frac{\omega_+^2-k^2}{2m_q^2}\right)\cdot\left(\frac{\omega_-^2-k^2}{2m_q^2}\right)\cdot\nonumber\\
{} & {} & \left[\displaystyle{e^{\displaystyle{-\beta(\omega_+-\omega_-)\tilde{\tau}}}}+\displaystyle{e^{\displaystyle{-\beta(\omega_+-\omega_-)(1-\tilde{\tau})}}}\right]+\left(\frac{1}{1+e^{-\beta\omega_-}}\right)^2\cdot\nonumber\\
{} & {} & \left.\cdot\left(\frac{\omega_-^2-k^2}{2m_q^2}\right)^2\left[\displaystyle{e^{\displaystyle{-2\beta\omega_-\tilde{\tau}}}}+\displaystyle{e^{\displaystyle{-2\beta\omega_-(1-\tilde{\tau})}}}\right]\right\}\cos^4(\alpha(k))+\nonumber\\
{} & {} & +\frac{2N_c}{\pi^2}\int_{\Lambda_1}^{\Lambda_2}k^2\,dk\left\{\frac{1}{1+e^{-\beta\epsilon_k}}\cdot\frac{1}{1+e^{-\beta\omega_+}}\cdot\right.\nonumber\\
{} & {} & \cdot\frac{\omega_+^2-k^2}{2m_q^2}\cdot\left[\displaystyle{e^{\displaystyle{-\beta(\epsilon_k+\omega_+)\tilde{\tau}}}}+\displaystyle{e^{\displaystyle{-\beta(\epsilon_k+\omega_+)(1-\tilde{\tau})}}}\right]+\nonumber\\
{} & {} & \frac{1}{1+e^{-\beta\epsilon_k}}\cdot\frac{1}{1+e^{\beta\omega_-}}\cdot\frac{\omega_-^2-k^2}{2m_q^2}\cdot\nonumber\\
{} & {} & \left.\cdot\left[\displaystyle{e^{\displaystyle{-\beta(\epsilon_k-\omega_-)\tilde{\tau}}}}+\displaystyle{e^{\displaystyle{-\beta(\epsilon_k-\omega_-)(1-\tilde{\tau})}}}\right]\right\}2\sin^2(\alpha(k))\cos^2(\alpha(k))+\nonumber\\
{} & {} & +\frac{2N_c}{\pi^2}\int_{\Lambda_1}^{\Lambda_2}k^2\,dk\left(\frac{1}{1+e^{-\beta\epsilon_k}}\right)^2\left[\displaystyle{e^{\displaystyle{-2\beta\epsilon_k\tilde{\tau}}}}+\displaystyle{e^{\displaystyle{-2\beta\epsilon_k(1-\tilde{\tau})}}}\right]\sin^4(\alpha(k))\;,\nonumber\\
\eeqa
\beqa
G_{(b)}^{\trm{pc}}(\tau) & = & \frac{2N_c}{\pi^2}\int_{\Lambda_1}^{\Lambda_2}k^2\,dk\int_{-k}^{k}d\omega_2\left\{\frac{1}{1+e^{-\beta\omega_+}}\cdot\frac{1}{1+e^{-\beta\omega_2}}\cdot\right.\nonumber\\
{} & {} & \cdot\frac{\omega_+^2-k^2}{m_q^2}\cdot\beta_+(\omega_2,k)\cdot\left[\displaystyle{e^{\displaystyle{-\beta(\omega_++\omega_2)\tilde{\tau}}}}+\displaystyle{e^{\displaystyle{-\beta(\omega_++\omega_2)(1-\tilde{\tau})}}}\right]+\nonumber\\
{} & {} & +\frac{1}{1+e^{-\beta\omega_-}}\cdot\frac{1}{1+e^{-\beta\omega_2}}\cdot\frac{\omega_-^2-k^2}{m_q^2}\cdot\beta_-(\omega_2,k)\cdot\nonumber\\
{} & {} & \left.\cdot\left[\displaystyle{e^{\displaystyle{-\beta(\omega_-+\omega_2)\tilde{\tau}}}}+\displaystyle{e^{\displaystyle{-\beta(\omega_-+\omega_2)(1-\tilde{\tau})}}}\right]\right\}\cos^4(\alpha(k))+\nonumber\\
{} & {} & +\frac{2N_c}{\pi^2}\int_{\Lambda_1}^{\Lambda_2}k^2\,dk\int_{-k}^{k}d\omega_2\frac{1}{1+e^{-\beta\epsilon_k}}\cdot\frac{1}{1+e^{-\beta\omega_2}}\cdot\beta_+(\omega_2,k)\cdot\nonumber\\
{} & {} & \cdot\left[\displaystyle{e^{\displaystyle{-\beta(\epsilon_k+\omega_2)\tilde{\tau}}}}+\displaystyle{e^{\displaystyle{-\beta(\epsilon_k+\omega_2)(1-\tilde{\tau})}}}\right]\cdot 2\cos^2(\alpha(k))\sin^2(\alpha(k))\nonumber\\
\eeqa 
and
\beqa
G_{(b)}^{\trm{cc}}(\tau) & = & \frac{2N_c}{\pi^2}\int_{\Lambda_1}^{\Lambda_2}k^2\,dk\int_{-k}^{k}d\omega_1\int_{-\omega_1}^k d\omega_2\frac{1}{1+e^{-\beta\omega_1}}\cdot\frac{1}{1+e^{-\beta\omega_2}}\cdot\nonumber\\
{} & {} & \cdot[\beta_+(\omega_1,k)\beta_+(\omega_2,k)+\beta_-(\omega_1,k)\beta_-(\omega_2,k)]\cdot\nonumber\\
{} & {} & \cdot\left[\displaystyle{e^{\displaystyle{-\beta(\omega_1+\omega_2)\tilde{\tau}}}}+\displaystyle{e^{\displaystyle{-\beta(\omega_1+\omega_2)(1-\tilde{\tau})}}}\right]\cdot\cos^4(\alpha(k))\;,
\eeqa
where $\alpha(k)$ has been defined in Eq. (\ref{eq:alphak}) and $\epsilon_k=\sqrt{k^2+m_\infty^2}$, being $m_\infty$ the NLA asymptotic mass.\\
Finally, the contribution coming from the hard momenta region reads:\\
\beq
G_{(c)}^{\trm{pp}}(\tau)=\frac{2N_c}{\pi^2}\int_{\Lambda_2}^{\infty}k^2\,dk\left(\frac{1}{1+e^{-\beta\epsilon_k}}\right)^2\left[\displaystyle{e^{\displaystyle{-2\beta\epsilon_k\tilde{\tau}}}}+\displaystyle{e^{\displaystyle{-2\beta\epsilon_k(1-\tilde{\tau})}}}\right]\;.
\eeq
Clearly one recovers the HTL approximation for $G(\tau,\bold{p}=\bold{0})$ by simply extending the domain of integration in Eqs. (\ref{eq:pphtl},\ref{eq:pchtl},\ref{eq:cchtl}) up to infinity and neglecting the other terms. 

\section{Fermion-antifermion loop}\label{c}
For the benefit of the reader we report some formulas useful for readily performing the sum over Matsubara frequencies inside a fermionic loop. Defining:
\beq\label{split}
\wtilde{\Delta}(i\omega_n,E_p)=\frac{1}{\omega_n^2+E_p^2}=\sum_{s=\pm1}\wtilde{\Delta}_{s}(i\omega_{n},E_p)=\sum_{s=\pm1}\frac{s}{2E_p}\frac{-1}{i\omega_n-sE_p}\; ,
\eeq
$\omega_{n}$ being odd, the following relationships hold \cite{lb}:
{\setlength\arraycolsep{1pt}
\beqa
T\sum_n\wtilde{\Delta}_{s_1}(i\omega_n,E_1)\wtilde{\Delta}_{s_2}(i\omega_l\!-i\omega_n,E_2) & = & \frac{-s_1s_2}{4E_1E_2}\frac{1-\tilde{n}(s_1E_1)-\tilde{n}(s_2E_2)}{i\omega_l-s_1E_1-s_2E_2}\nonumber\\
\label{sum1}\\
T\sum_n\omega_n\wtilde{\Delta}_{s_1}(i\omega_n,E_1)\wtilde{\Delta}_{s2}(i\omega_l\!-i\omega_n,E_2) & = & \frac{is_2}{4E_2}\frac{1-\tilde{n}(s_1E_1)-\tilde{n}(s_2E_2)}{i\omega_l-s_1E_1-s_2E_2}\; ,\nonumber\\ \label{sum2}
\eeqa} 
where $\omega_l$ is even and $\tilde{n}(\omega)$ is a Fermi distribution. We remind the useful identity:
\beq
\tilde{n}(-\omega)=1-\tilde{n}(\omega)\;.
\eeq

\end{document}